\documentclass[fleqn,usenatbib]{mnras}

\usepackage[T1]{fontenc}
\usepackage{ae,aecompl}

\usepackage{graphicx}	
\usepackage{amsmath}	
\usepackage{amssymb}	
\usepackage{bm}
\usepackage{arydshln}   
\usepackage{multirow}
\usepackage[dvipsnames]{xcolor}

\usepackage[normalem]{ulem}  

\usepackage{hyperref} 
\usepackage{tikz}
\definecolor{lime}{HTML}{A6CE39}

\DeclareRobustCommand{\orcidicon}{
	\begin{tikzpicture}
	\draw[lime, fill=lime] (0,0) 
	circle [radius=0.16] 
	node[white] {{\fontfamily{qag}\selectfont \tiny ID}};
	\draw[white, fill=white] (-0.0625,0.095) 
	circle [radius=0.007];
	\end{tikzpicture}
	\hspace{-2mm}
}
\foreach \x in {A}{%
	\expandafter\xdef\csname orcid\x\endcsname{\noexpand\href{https://orcid.org/\csname orcidauthor\x\endcsname}{\noexpand\orcidicon}}
}

\graphicspath{{plot/}}

\usepackage{xspace}
\newcommand{\gadgetx}{\textsc{Gadget-X}\xspace}
\newcommand{\ahf}{\textsc{AHF}\xspace}
\newcommand{\oPDF}{\textsc{oPDF}\xspace}

\newcommand{\ud}{\mathrm{d}}

\usepackage{soul}

\usepackage{newtxtext,newtxmath}

\title[cluster dynamical tracers]{What to expect from dynamical modelling of cluster haloes I. The information content of different dynamical tracers}

\author[Li et al.]{\parbox{\textwidth}{
Qingyang Li\orcidA{},$^{1,2}$\thanks{qingyli@sjtu.edu.cn}
Jiaxin Han,$^{1,2}$\thanks{jiaxin.han@sjtu.edu.cn}
Wenting Wang,$^{1,2}$
Weiguang Cui,$^{3}$
Zhaozhou Li,$^{1,2}$
Xiaohu Yang$^{1,2,4}$
}
\vspace{0.4cm}
\\
\parbox{\textwidth}{
$^1$Department of Astronomy, School of
  Physics and Astronomy, Shanghai Jiao Tong University, Shanghai 200240, China\\
$^2$Shanghai Key Laboratory for Particle Physics and Cosmology, Shanghai 200240, China\\
$^3$Institute for Astronomy, University of Edinburgh, Royal Observatory, Edinburgh EH9 3HJ, United Kingdom\\
$^4$Tsung-Dao Lee Institute, Shanghai Jiao Tong University, Shanghai 200240, China\\
}}

\date{Accepted 2021 May 31. Received 2021 May 24; in original form 2021 April 25}

\pubyear{2021}

\begin{document}
\label{firstpage}
\pagerange{\pageref{firstpage}--\pageref{lastpage}}
\maketitle

\begin{abstract}
Using hydrodynamical simulations, we study how well the underlying gravitational potential of a galaxy cluster can be modelled dynamically with different types of tracers. In order to segregate different systematics and the effects of varying estimator performances, we first focus on applying a generic minimal assumption method (\oPDF) to model the simulated haloes using the full 6-D phasespace information.
We show that the halo mass and concentration can be recovered in an ensemble unbiased way, with a stochastic bias that varies from halo to halo, mostly reflecting deviations from steady state in the tracer distribution. 
The typical systematic uncertainty is $\sim 0.17$ dex in the virial mass and $\sim 0.17$ dex in the concentration as well when dark matter particles are used as tracers. 
The dynamical state of satellite galaxies are close to that of dark matter particles, while intracluster stars are less in a steady state, resulting in a $\sim$ 0.26 dex systematic uncertainty in mass. Compared with galactic haloes hosting Milky-Way-like galaxies, cluster haloes show a larger stochastic bias in the recovered mass profiles. We also test the accuracy of using intracluster gas as a dynamical tracer modelled through a generalised hydrostatic equilibrium equation, and find a comparable systematic uncertainty in the estimated mass to that using dark matter. Lastly, we demonstrate that our conclusions are largely applicable to other steady-state dynamical models including the spherical Jeans equation, by quantitatively segregating their statistical efficiencies and robustness to systematics. We also estimate the limiting number of tracers that leads to the systematics-dominated regime in each case.
\end{abstract}



\section{Introduction} \label{sec:intro}
Galaxy clusters are the largest gravitationally bound systems in the Universe and are unique laboratories for studying cosmology and galaxy formation. Accurate and robust estimation of the mass content of clusters is crucial for such studies, especially in the era of precision cosmology. Besides gravitational lensing \citep[e.g.,][]{Mandelbaum2006,Hilbert2010,Richard2010,Han2015}, dynamical modelling \citep[e.g.,][]{Biviano2006,Lau2009,Rasia2012,Mamon2013,Old2014,Wang2015,Henson2017,Armitage2019} is perhaps the most popular approach to model the underlying potential and measure the total mass of galaxy clusters. 

A dynamical model generally makes use of the phase-space distribution of tracer particles to infer the underlying gravitational potential, as the motions of tracers are governed by the gravity of the host halo. A large family of methods have been developed in the past, including the Jeans equation~\citep[e.g.,][]{Binney1987}, the virial theorem~\citep[e.g.,][]{Heisler1985}, the caustic method~\citep[e.g.,][]{Diaferio1997}, the distribution function method~\citep[e.g.,][]{Li2019}, and the hydrostatic equilibrium equation~\citep[e.g.,][]{Rasia2004} in a general categorization. In addition, machine learning has also become one popular approach to measure the dynamical masses for galaxy clusters in recent years~\citep[e.g.,][]{Ntampaka2015,Ramanah2020,Ramanah2021}. Many works have been devoted to testing and understanding the various dynamical methods in both simulations and observations~\citep[e.g.,][]{Carlberg1997,Girardi1998,Biviano2003,Rines2003,Lemze2009,Serra2011,Alpaslan2012,Biviano2013,Munari2014,Maughan2016,Foex2017,Capasso2019,Armitage2019,Lovisari2020}, finding different performances of different methods.

Because bright satellite galaxies are relatively easy to observe in phase space than individual stars, they have been the primary dynamical tracer objects for modelling the underlying gravitational potential in clusters~\citep[e.g.,][]{Biviano2003,Old2015,Old2018,Wojtak2018}. \citet{Saro2013} investigated the selection effects on the estimation of galaxy cluster masses using a sample of red-sequence and spectroscopic galaxies selected from a simulated catalogue. \citet{Armitage2019} tested a few traditional dynamical modelling methods including the spherical Jeans equation, the virial theorem and the caustic method, using satellite galaxies as tracers from the C-EAGLE simulation. In both  3-dimensional (hereafter 3D) and projected cases, they found similar performances of the three methods. However, \citet{Old2015} studied the scatter and bias of 25 mass estimation techniques using cluster galaxies and found $0.18-1.08$ dex rms errors in the recovered virial mass for different methods.

Apart from satellite galaxies, hot gas in clusters can be used as dynamical tracers as well. The X-ray emission from the hot intracluster medium (ICM) provides estimates of the density and temperature of the ICM. Under the assumption of hydrostatic equilibrium, these quantities can be used to infer the total mass profile of the host cluster halo~\citep{Cowie1987,Diaferio2005,Vikhlinin2006,Rines2016,Maughan2016,Foex2017,Lovisari2020}. However, there could be systematic differences between such hydrostatic mass estimates and other dynamical masses. For example, \citet{Maughan2016} found that the hydrostatic-to-caustic mass ratio is about 1.2 at $R_{500}$\footnote{ The notation $R_{500}$ is defined as the radius where the mean density is 500 times the critical density of the universe at the cluster redshift}.

More detailed analysis revealed that dynamical substructures, which can be detected with the Dressler-Shectman test \citep{Dressler1988}, can have significant influences on the dynamical mass estimates~\citep[e.g.,][]{Biviano2006,Foex2017,Old2018,Armitage2019}. In observations, \citet{Foex2017} showed that the removal of substructures can result in $\sim 15\%$ smaller dynamical masses in general.  On the other hand, using mock galaxy catalogues, \citet{Old2018} found that the scatter in the dynamical mass estimates of clusters with dynamical substructures is very close to that without substructures.

Using data from the wide-field VIMOS spectroscope and XMM-Newton observations, \citet{Foex2017} found that the Jeans, caustic and virial methods with substructures, can lead to $\sim20$\%, $\sim30$\% and $\sim50$\% higher mass estimates than the mass estimated from X-ray observations of the hot ICM, respectively. In fact, the observed X-ray emission can be affected by many physical processes in clusters, such as gas accretion, active galactic nuclei (AGN) feedback and the presence of substructures, which violate the hydrostatic equilibrium assumption. These effects have been widely studied in numerical simulations~\citep[e.g.,][]{Rasia2006,Lau2009,Rasia2012,Nelson2014,Henson2017,Ansarifard2020, Gianfagna2021}, showing about 10 $-$ 40\% systematic errors in the mass estimates.

Many of these previous studies, however, have focused on testing the performance of specific dynamical methods that are commonly used in observations. While such tests can provide valuable benchmarks for the applications of the corresponding methods, it is still not easy to get a concise and consistent understanding of the most important systematics, due to the various peculiar assumptions associated with different models. On the other hand, a few assumptions, such as the steady-state and the spherical symmetric assumptions, are commonly adopted by almost all of the dynamical modelling methods. Understanding the systematics arising from these common assumptions can help us to understand the minimum level of systematics with which one can expect from models involving these assumptions. To this end, in this work we carry out a comprehensive investigation of common systematics by applying a generic minimal assumption dynamical model to simulations of galaxy clusters. 
We use a large sample of galaxy clusters from a set of high resolution hydrodynamical simulations for this study. Various types of tracers are adopted in our analysis, including dark matter (hereafter DM) particles, halo stars, satellite galaxies and intra-cluster gas. Although it is impossible or difficult to use DM or individual halo stars as dynamical tracers in distant cluster observations, tests using these tracers in simulations can nonetheless provide a comprehensive view of the dynamical state of clusters. In fact, a few previous studies have also attempted using DM particles in numerical simulations as tracers. For example, \citet{Marini2021} studied the pseudo-entropy profiles using DM, stars and subhaloes as tracers in simulated galaxy clusters. \citet{Biviano2006} used both DM and satellite galaxies as tracers to estimate the mass of clusters in simulations with the virial theorem. \citet{Wojtak2009} used DM as tracers to test the distribution function method of constraining the mass and anisotropy profiles of galaxy clusters.

To isolate systematics due to any peculiar assumptions associated with a specific dynamical method, we at first use a generic dynamical method that has minimal model assumptions and test the method using these different types of tracers. This method, called the orbital Probability Distribution Function~\citep[\oPDF;][]{Han2016a}, has been used to study the dynamical state of various tracers in Milky Way (MW) like haloes~\citep{Han2016b,Wang2017,Han2020}. It has been found that MW haloes are only approximately in a steady state, resulting in an \emph{irreducible} amount of bias, which is determined by the dynamical state of the corresponding dynamical tracers in the haloes. While satellite galaxies show a dynamical state close to that of DM~\citep{Han2020}, halo stars deviate more from equilibrium, leading to a factor of $\sim2-3$ systematic bias in the mass estimates~\citep{Wang2017}. The conclusions are further validated using the spherical Jeans equation (hereafter SJE) modelling in a separate work~\citep{Wang2018}. In this study, we show that similar conclusions hold for cluster haloes, but the amount of the overall bias level is larger. This can be understood as cluster haloes form later than galaxy size haloes, leading to stronger deviations from equilibrium. 

We also complement the \oPDF results by comparing them to those based on the hydrostatic equilibrium modelling of hot gas in clusters as well as those based on the SJE, to understand the general implications of our results especially in terms of quantifying the efficiency and robustness of each estimator.

The current version of \oPDF is based on the full 6-dimensional phase-space information of tracer objects. However, in observations we can only observe the line-of-sight velocity component of satellite galaxies, because it is very difficult to measure the proper motions or tangential velocities for satellite galaxies in distant clusters. Nevertheless, as our first attempt, in this paper we apply the full phase-space \oPDF method to galaxy clusters, using the 6-dimensional position and velocity information for various types of tracer particles in numerical simulations. This helps us to avoid extra assumptions due to incomplete data, while focusing on investigating the dynamical state of different types of tracers. We leave the application to incomplete phase-space data to future studies.

This paper is organised as follows. Section~\ref{sec:data} gives a brief introduction to the simulation data used. The dynamical modelling methods are summarised in Section~\ref{sec:method}. The overall performances of dynamical modelling in cluster haloes using different dynamical tracers as described above are presented in Section~\ref{sec:results}. We will also show comparisons with MW-like systems, which we refer to as galactic haloes throughout the paper. In Section~\ref{sec:sys}, we discuss potential systematic errors caused by the spherical symmetry assumption, radial selections and baryonic effects. In Section~\ref{sec:SJE}, we demonstrate the general implication of our results by further testing the SJE on the same tracer samples. We conclude in the end (Section~\ref{sec:summary}).

\section{Data} \label{sec:data}
We use a sample of 324 simulated galaxy clusters from the Three Hundred Project\footnote{\url{https://the300-project.org}}~\citep{Cui2018} for our analysis. These clusters are selected from the parent $N$-body simulation of MultiDark Planck 2~\citep[MDPL2;][]{Klypin2016}\footnote{\url{https://www.cosmosim.org/cms/simulations/mdpl2}}, which was run with $3840^3$ dark matter particles of mass $1.5\ \times$ $10^9\ h^{-1}\mathrm{M_{\odot}}$ and in a periodic cube of comoving length 1.48 Gpc. Each selected cluster zooms into a region of 15 $h^{-1}\mathrm{Mpc}$ in radius centred on one of the 324 most massive clusters at $z=0$ in MDPL2, as identified by the \textsc{rockstar}~\citep{Behroozi2012} halo finder\footnote{\url{https://bitbucket.org/gfcstanford/rockstar}}. The particle masses in the high resolution region are $m_{\rm DM}\simeq 12.7 \times 10^8\ h^{-1}\mathrm{M_{\odot}}$ and $m_{\rm gas} \simeq 2.36\ \times 10^8\ h^{-1}\mathrm{M_{\odot}}$.  These cluster regions are re-simulated with different hydrodynamical codes: \textsc{GADGET-MUSIC}, \gadgetx, {\sc GIZMO-SIMBA} (Cui et al. in prep.) using the zoom-in technique under the same Planck cosmology as the parent simulation~\citep{Planck2016}. These simulated clusters have been used for different studies, for example, environment effect \citep{WangYang2018}, cluster profiles \citep{Mostoghiu2019,Li2020,Baxter2021}, splash-back galaxies \citep{Arthur2019,Haggar2020,Knebe2020}, cluster dynamical state \citep{DeLuca2021,Capalbo2021}, filament structures \citep{Kuchner2020,Rost2021,Kuchner2021} and lensing studies \citep{Vega-Ferrero2021}.

In this work, our default galaxy cluster samples are generated with \gadgetx\ which is based on the gravity solver of the {\sc GADGET3} Tree-PM code (an updated version of the {\sc GADGET2} code; \citealt{Springel2005}), using smoothed particle hydrodynamics (SPH) to follow the evolution of the gas component. The SPH scheme \citep{Beck2016} is improved with artificial thermal diffusion, time-dependent artificial viscosity, a high-order Wendland C4 interpolating kernel and a weak-up scheme. Stellar evolution and metal enrichment in \gadgetx\ \citep[see][for the original formulation]{Tornatore2007} consider mass-dependent lifetimes of stars \citep{Padovani1993}, the production and evolution of 15 different elements coming from SNIa, SNII and AGB stars with metallicity-dependent radiative cooling \citep{Wiersma2009}. The stellar feedback model is adopted from \citet{Springel2003} with a wind velocity of $400\ \rm km\ s^{-1}$. The \gadgetx\ model also includes black hole (BH) growth and an implementation of AGN feedback \citep{Steinborn2015}. Lastly, \gadgetx\ adopts the Chabrier initial mass function \citep[IMF,][]{Chabrier2003}. 

The haloes in these re-simulations are identified by the Amiga Halo Finder~\citep[\ahf;][]{Knollmann2009} with an overdensity of 200 $\times\  \rm \rho_{crit}$, where $\rm \rho_{crit}$ is the critical density of the Universe at the corresponding redshift. In our analysis, we only use the central clusters as our sample and consider the virial mass $M_{200}$\footnote{The virial mass, $M_{200}$, is defined as the total mass enclosed in a radius, $r_{200}$, within which the average matter density is 200 times the mean critical density of the universe.} identified by \ahf as the true value for each halo. The true halo concentration is defined as $r_{200}/r_\mathrm{s}$ from \citet{Cui2018}, where the parameter $r_\mathrm{s}$ is obtained by fitting a Navarro-Frenk-White~\citep[NFW;][]{NFW} model to the radial total mass density profile at $r>0.01r_{200}$. \ahf also outputs subhaloes within each cluster. Particles within subhaloes are subject to the gravitational influence of the subhalo itself, which are not accounted for by our dynamical models. As a result, by default we exclude DM, star and gas particles contained in any subhaloes (or satellites) from our tracer sample\footnote{When satellites are used as dynamical tracers, each satellite is treated as one tracer object, i.e., DM, star or gas particles belonging to this subhalo or satellite are not used as individual tracers.}, following \citet{Han2016b} and \citet{Wang2017}. The tracer particles in subhaloes are only included when explicitly examining the effect of subhalo contamination. Lastly, when satellite galaxies are used as tracers, a stellar mass cut of $10^9\ \rm  M_{\odot}$ is applied to the satellite galaxy sample to mimic the stellar mass limit of upcoming surveys such as Euclid~\citep{Laureijs2011}, following \citet{Armitage2019}. 

Besides the clusters from the hydrodynamical simulations, we will also analyse the corresponding cluster haloes identified by \ahf from the parent MDPL2 $N$-body simulation, using dark matter particles as tracers. The results will be compared with those based on the \gadgetx haloes, to learn about the effect of baryonic physics on dynamical models.

\section{Method} \label{sec:method}
As introduced in Section~\ref{sec:intro}, different dynamical modelling methods usually involve their own peculiar assumptions, which may lead to systematic errors. To achieve a general assessment of the dynamical state of cluster haloes and the associated limiting accuracy in dynamical modelling results, we will use a generic dynamical model that involves only the most commonly used assumptions. The \oPDF method~\citep{Han2016a} adopts minimal assumptions -- the steady-state and spherical symmetry assumptions, which will be presented in more details below. This method is applicable to collisionless tracers such as DM and star particles as well as satellite galaxies. For completeness, we also apply the commonly used hydrostatic equilibrium equation to gas particles in the simulation in Section~\ref{sec:SJE}. The results from the hydrostatic equilibrium model will be compared with the results based on \oPDF using collisionless tracers. 

Both \oPDF and the hydrostatic equilibrium model can be applied to recover the underlying potential non-parametrically. In this work, however, we focus on modelling the parametrized potential profile of the halo. More specifically, the potential profile, $\Phi(r)$, is parametrized by generalizing the true potential to a template with the formula $\Phi(r) = A \Phi_{\rm true}(Br)$, where the true profile $\Phi_{\rm true}(r)$ is extracted from the simulation, and $A$ and $B$ are free parameters which can be converted to mass and concentration of clusters following \citet{Han2016a}. Our results are thus presented in terms of the best-fitting halo mass and concentration parameters ($M$ and $c$) versus their true values. Here $M$ denotes the virial mass, $M_{200}$, as defined above, and concentration is defined as $c=r_{200}/r_{\rm s}$, where $r_{200}$ is the virial radius and $r_{\rm s}$ is the radius at which the density profile has a logarithmic slope of $-2$. Such a parametrization is compatible with the commonly adopted NFW profile of haloes. The advantage of using such potential templates extracted from the true potential and mass profiles is that it avoids systematics caused by deviations from the NFW form for each individual halo. Hence with the potential template models to parametrize the underlying potential, we can focus on investigating the remaining systemics. The sources of remaining systematics of \oPDF are deviations from the steady state and from spherical symmetry. In this paper, we mainly focus on discussions of the steady-state assumption. In a follow-up study, we will provide more detailed investigations on different proxies of the dynamical state and the driving variables of the systematics including halo shape. 

We should bear in mind that such a template parametrization is not applicable to real data, for which we do not know the true potential profile beforehand. To assess the importance of a proper parametrization of the underlying potential, we also carry out tests using \oPDF with an NFW profile, and compare the results with those based on the templates. 

\subsection{The orbital Probability Density Function (\oPDF) method}
\oPDF \footnote{Code available at \url{https://github.com/Kambrian/oPDF}} is a freeform distribution function method that works by fitting a data-driven distribution function to the tracer sample to infer the underlying potential. More specifically, the method first predicts a steady-state spatial distribution of tracer particles from their observed phase space coordinates once an underlying potential is assumed. The predicted spatial distribution is then compared against the observed tracer distribution to infer the correct potential.
The prediction of the steady-state distribution relies on the orbital Probability Distribution Function, which states that if a system is in a steady state, then along a given orbit the probability of observing a particle near any position (labelled by the parameter $\lambda$) is proportional to the time that the particle spends around that position, i.e.,
\begin{equation}
    \frac{\ud P(\lambda|{\rm orbit})}{\ud\lambda} \propto \frac{\ud t(\lambda|{\rm orbit})}{\ud\lambda}.
\end{equation} This conditional distribution can be derived from the time-independent collisionless Boltzmann equation, and is equivalent to the Jeans theorem~\citep{Han2016a}. 

When applied to a spherical system, the orbital distribution can be expressed via a radial coordinate $r$ as follows:
\begin{equation}
    \ud P(r|E,L)= \frac{\ud t}{\int \ud t} = \frac{1}{T}\frac{\ud r}{|v_r|},
\label{eq_oPDF}
\end{equation}
where $E$, $L$ and $T$ are the binding energy, angular momentum and the period of the orbit respectively. The radial velocity at any radius in the conservative central force field is
\begin{equation} 
    v_r = \sqrt{2\Phi (r) - 2E - L^2/r^2},
\end{equation}
where $\Phi (r)$ is the model potential. 

For each tracer particle $i$ with observed position $\bm{r}_i$ and velocity $\bm{v}_i$, the orbital parameters can be obtained as $E_i=-(\bm{v}^2_i/2 + \Phi (r_i))$ and $L_i=| \bm{r}_i \times \bm{v}_i|$ once a model potential $\Phi(r)$ is adopted. The overall steady-state radial distribution of all the particles $N$ is then predicted combining the oPDF of each particle as
\begin{equation}
    P(r)=\frac{1}{N} {\textstyle\sum_i} P(r|E_i,L_i).\label{eq:P_r}
\end{equation}

With this predicted distribution function, the underlying potential can be inferred by matching the predicted distribution with the observed distribution in a statistical framework. Here, we take the binned likelihood approach for the inference. If the data is separated into $m$ bins, the number of particles in the $j$-th bin is expected to be
\begin{equation}
    \hat{n}_j = N \int_{r_{l,j}}^{r_{u,j}} \frac{\ud P(r)}{\ud r} \ud r,
\end{equation}
where $r_{l,j}$ and $r_{u,j}$ are the lower and upper bounds of the bin. The likelihood of observing the data with the model distribution of Equation~\eqref{eq:P_r} can be written as 
\begin{equation}
    \begin{split}
        \mathcal{L} &=\bm{\prod}_{j=1}^m \hat{n}_j^{n_j} \exp(-\hat{n}_j) \\
        &=\exp(-N) \bm{\prod}_{j=1}^m \hat{n}_j^{n_j},
    \end{split}
\end{equation}
where $n_j$ means the observed number of particles in the $j$-th bin. The true potential can be inferred by searching for a potential that maximizes this likelihood.

\section{Overall dynamical state and the irreducible bias of steady-state models} \label{sec:results}
In this section, we study the overall dynamical state of galaxy clusters using different tracers, and compare the results against each other and with those of the galactic haloes hosting MW-mass galaxies. As introduced before, we use the free-form orbital probability distribution function method, \oPDF, to model the phase-space distribution of DM, halo stars and satellite galaxies, by adopting a parametrized template potential profile, generalised from the true underlying potential. For the intracluster gas, the generalised hydrostatic equilibrium equation will be used for dynamical modelling.

\subsection{Dark matter}
\begin{figure}
    \centering
    \includegraphics[width=0.5\textwidth]{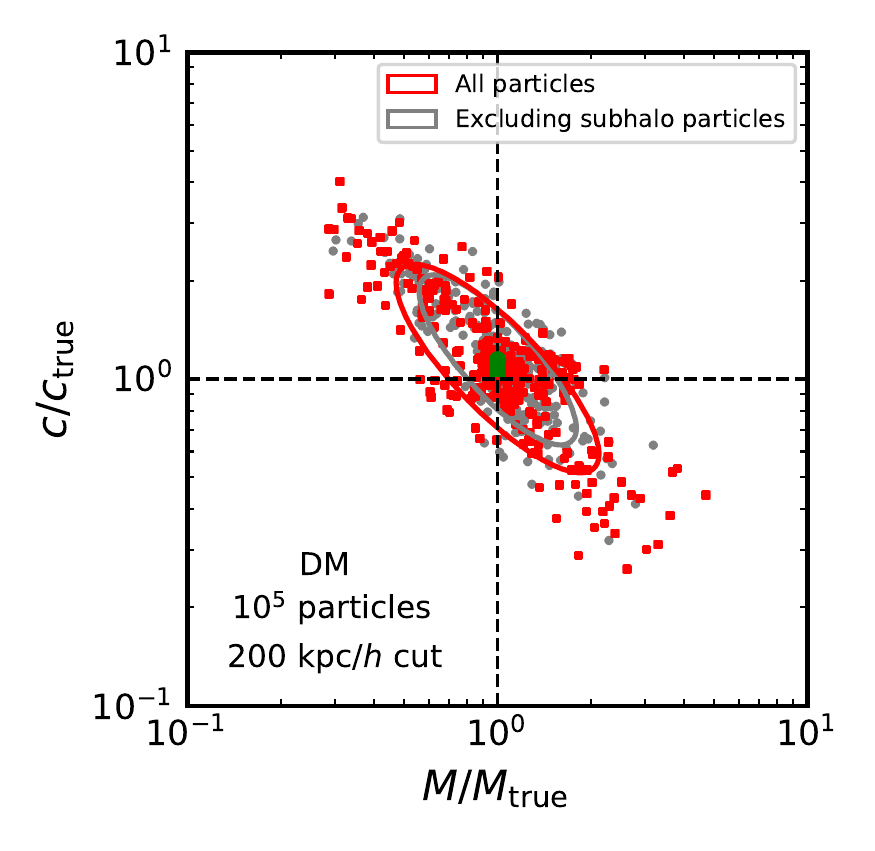}
    \caption{The best-fitting mass and concentration parameters of cluster haloes using the \oPDF method with DM particles as tracers, normalised by their true values. Each point shows the best-fitting parameters for one cluster halo. The red squares and grey circles show results using tracers before and after removing tracer particles in substructures, respectively.
    In either case, $10^5$ randomly selected DM particles from an inner radius of 200 $h^{-1}\rm kpc$ out to the virial radius are used for each halo. 
    The ellipses show the $1\sigma$ scatter of both samples,
    while the bold green symbols indicate the mean values, which are almost ensemble unbiased.
    }
    \label{fig:DMsub}
\end{figure}

The distribution of the best-fitting halo parameters using DM particles as tracers are shown in Fig.~\ref{fig:DMsub}. For each cluster, a sample of $10^5$ randomly selected particles in the radial range of $200\ h^{-1}$kpc $< r < r_{200}$ are used as tracers. The inner cut is adopted to eliminate the effect caused by inaccuracies in the determination of halo centres as well as to avoid uncertainties associated with the baryonic treatment in the simulation, which we will further discuss in section~\ref{sec:BE}. To eliminate catastrophic failures in the fits, we remove clusters whose fitted parameters lie outside the 3-$\sigma$ confidence region of the sample distribution. This is done iteratively until the remaining number of clusters converge. The same operation is applied to all the other similar plots in this paper, resulting in about 310 clusters on average.

Substructures in dark matter haloes can lead to deviations from the steady-state assumption. To test the influence of such substructures, we carry out a comparison by applying the \oPDF method to DM particles both including and excluding subhalo particles. For the latter, we remove all the particles bound to subhaloes from the tracer sample. Note that even the unbounded particles near the subhalo are still suffering from its potential, thus leads to deviations from the steady-state assumption.

As shown in Fig.~\ref{fig:DMsub}, the best-fitting results using both samples of tracers show large amounts of scatter around the true parameter values, reflecting deviations from the model assumptions. With particles in substructures removed, the total scatter in ${\log M/M_{\rm true}}$(${\log c/c_{\rm true}}$) is reduced from 0.22(0.21) dex to 0.17(0.17) dex. This is consistent with our expectation that particles in substructures can violate the steady-state assumption by perturbing the phase-space structure of tracers. 
In contrast to the above significant difference for cluster haloes, \citet{Han2016b} reported a difference of $\sim 1\%$ in the best-fitting halo parameters for galactic haloes with and without tracer particles in substructures. This can be understood as galactic haloes typically form earlier in the hierarchical universe and thus have more time to relax dynamically. The fraction of the total mass in subhaloes or satellites of galactic haloes are also smaller compared with that in cluster haloes~\citep{Han2015}. In the following, we will only use particles that do not belong to any subhaloes as tracers, and call these particles as smooth halo particles.

\begin{figure}
    \centering
    \includegraphics[width=0.5\textwidth]{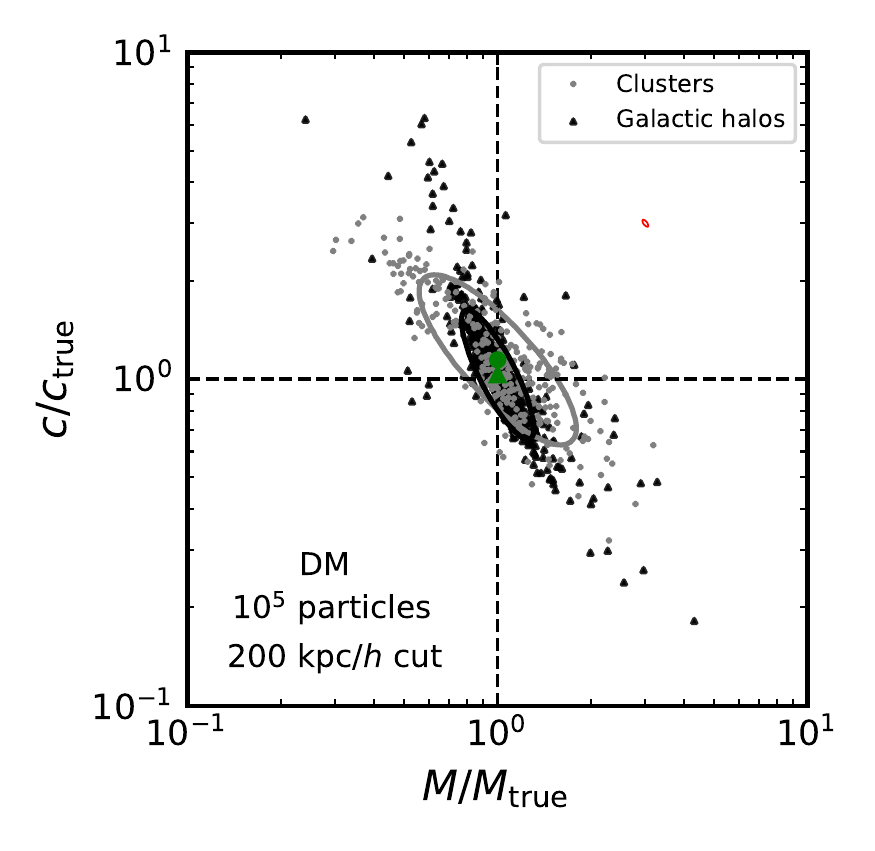}
    \caption{The best-fitting mass and concentration parameters of cluster and galactic haloes using DM particles as tracers. Only smooth halo particles (i.e., not within subhaloes) are used for the fitting. The grey points are fits to cluster haloes with an inner radius cut of $200\ h^{-1}{\rm kpc}$, while the black points are fits to galactic haloes taken from \citet{Wang2017} with an inner radius cut of 20~kpc. The grey and black ellipses indicate the 1$\sigma$ scatters for cluster and galactic haloes, respectively. The green triangle and circle are the mean values of the galactic and cluster fits respectively. The red ellipse in the upper right corner shows the typical statistical error of a single cluster fitting (translated to the current location for clarity), as inferred from the likelihood function. 
    }
    \label{fig:overall_DM}
\end{figure}

In Fig.~\ref{fig:overall_DM}, we directly compare the best-fitting results of cluster haloes with those of galactic haloes from \cite{Wang2017}. Both results are based on the \oPDF method using smooth halo particles. \citet{Wang2017} analysed a sample of $\sim1200$ Milky-Way like isolated or binary haloes, selected from the cosmological Millennium-II 
$N$-body simulation \citep{Boylan-Kolchin2009}, spanning a mass range of $0.5 \times 10^{12}\ \mathrm{M_{\odot}} < M_{200} < 2.5 \times 10^{12}\ \mathrm{M_{\odot}}$. The average number of DM particles used in each galactic halo is also about $10^5$, which is comparable to our sample size. The large number of tracers in each halo lead to a very small formal statistical error inferred from the likelihood profile, as shown by the red ellipse in the upper right corner of the figure for a typical cluster halo. If the model is an accurate description of the data, then the statistical error should match the total scatter in the fitted parameters. Given this small statistical uncertainty, the observed deviations in the best-fitting parameters from their true values are thus primarily due to systematic biases in the fitting. The mean value of ${M/M_{\rm true}}$ (${\log M/M_{\rm true}}$) for clusters is about 1.08 (0.00) with a dispersion of 0.41 (0.17). Even after excluding tracers in substructures, cluster haloes still show a significantly larger scatter than galactic haloes in Fig.~\ref{fig:overall_DM}. The readers can refer to Table~\ref{tab:all} for more detailed statistics about the mean biases and dispersions. The dispersions from our analysis are comparable to the results of \citet{Mamon2013} using the MAMPOSSt method applied to DM tracers in projection, who found dispersions of $\sim 30\%$ in mass and $\sim 48\%$ in concentration.


As what have been found in \citet{Wang2017}, these systematic biases are stochastic and largely ensemble unbiased. At the same time, the biases in the best-fitting mass and concentration parameters are anti-correlated along the same direction for both cluster and galactic haloes. This also coincide with the correlation direction in the statistical uncertainties shown by the red ellipse. It can be explained in the following way. The tracer particles used for dynamical modelling are correlated with each other in phase space. For example, particles stripped from the same progenitor subhalo or satellite share similar orbits with similar phase-space coordinates initially, and only gradually disperse away from each other to become more phase-mixed. As a result, strong phase correlations are expected among particles stripped from the same progenitor, resulting in a much smaller effective number of phase-independent particles. 
In other words, the statistical error ellipse underestimates the total uncertainty due to the correlations in the phase-space coordinates of tracers, leading to the observed stochastic biases. However, one should bear in mind that these stochastic biases are still a source of systematic uncertainties, as phase correlations violate the steady-state assumption, i.e., the phase-space density will evolve over time due to the existence of correlated structures. Consistent with \citet{Wang2017}, we have verified that increasing the number of tracer particles further reduces the statistical uncertainty. However, the observed total scatter barely changes as it is dominated by the stochastic bias due to the phase correlations. With this interpretation, the larger scatter of cluster haloes compared to galactic haloes, reflects a stronger deviation from a complete phase mixing or a steady state, that is consistent with the late formation time of galaxy clusters which have less time to erase the phase correlations.

Following \citet{Wang2017}, if we interpret the correlation between the statistical uncertainty and the total scatter as the phase correlation among tracer particles, we can define an effective number of the phase-independent particles as the tracer sample size. Using the effective number of particles will lead to a similar formal statistical error as its total scatter in the best-fitting parameters, i.e.,
\begin{equation} \label{eq:Neff}
    \Sigma_{\rm tot} = \frac{N_{\rm tracer}}{N_{\rm eff}}\Sigma_{\rm sta},
\end{equation}
where $N_{\rm eff}$ and $N_{\rm tracer}$ are the effective and actual tracer sample size, and $\Sigma_{\rm tot}$ and $\Sigma_{\rm sta}$ are the total and statistical covariance matrices.
To eliminate the correlation between the actual and effective number of tracer particles, we define an intrinsic number of tracer particles, $N_{\rm int}$, which is only responsible for the systematic uncertainties, $\Sigma_{\rm sys}$, through \citep{Wang2018},
\begin{equation} \label{eq:Nint}
    \begin{split}
    \Sigma_{\rm tot} &= \Sigma_{\rm sys} + \Sigma_{\rm sta} \\
    &= \frac{N_{\rm tracer}}{N_{\rm int}}\Sigma_{\rm sta} + \Sigma_{\rm sta}. \\
    \end{split}
\end{equation}
It then follows that
\begin{equation}
1/N_{\rm eff}=1/N_{\rm int}+1/N_{\rm tracer}.\label{eq:Ndecomp}
\end{equation}
In the systematics dominated regime, $\Sigma_{\rm tot}\approx\Sigma_{\rm sys}$ and $N_{\rm eff}\approx N_{\rm int}\ll N_{\rm tracer}$. In the statistics dominated regime, however, $N_{\rm eff}\approx N_{\rm tracer}\ll N_{\rm int}$.

In Fig.~\ref{fig:overall_DM}, the size of the grey ellipse is about 23 times larger than the size of the red ellipse. This means $N_{\rm eff}$ is roughly $1/23^2$ times the actual number of tracers. As we use $10^5$ tracer particles for each cluster, the effective number of phase-independent particles is $N_{\rm eff}\approx N_{\rm int} \sim 10^5/23^2 \approx 2\times10^2$. Note, however, the $N_{\rm eff}$ and $N_{\rm int}$ estimated in this way should be regarded as lower limits as we have ignored other sources of systematics, especially the violations to the spherical symmetry, which we will discuss further in Section~\ref{sec:sph}. Nevertheless, it is much smaller than the number of $N_{\rm eff}\sim 10^3$ for galactic haloes according to \citet{Wang2017}, reflecting a higher phase correlations among cluster particles. This also means that the cluster mass estimates using more than $2\times10^2$ DM particles as tracers will start to enter the systematics-dominated regime according to Equation~\eqref{eq:Nint}.

The separation between systematic and statistical errors has also been investigated in some previous works~\citep[e.g.,][]{Saro2013,Wojtak2018,Ramanah2020}.
\citet{Ramanah2020} used a formula given by \citet{Wojtak2018} to split the total error and compared the precision of their neural flow mass estimator with another seventeen dynamical methods investigated in \citet{Old2015}. Their results \citep[Fig.6 in][]{Ramanah2020} showed that the systematic errors for most methods mainly cluster around $0.10-0.20$ dex, while statistical errors concentrate on about $0.05-0.15$ dex. Our error estimation based on Eq.~\eqref{eq:Nint} gives a similar performance with $\sigma_{\rm sys} \sim 0.13$ and $\sigma_{\rm sta} \sim 0.07$ when using satellite galaxies as tracers (see details in Table.~\ref{tab:all} and Section.~\ref{sec:sat}). We will investigate the levels of systematics associated with different estimators in more detail in section~\ref{sec:robustness}.

\subsection{Halo stars} 

\begin{figure}
    \centering
    \includegraphics[width=0.5\textwidth]{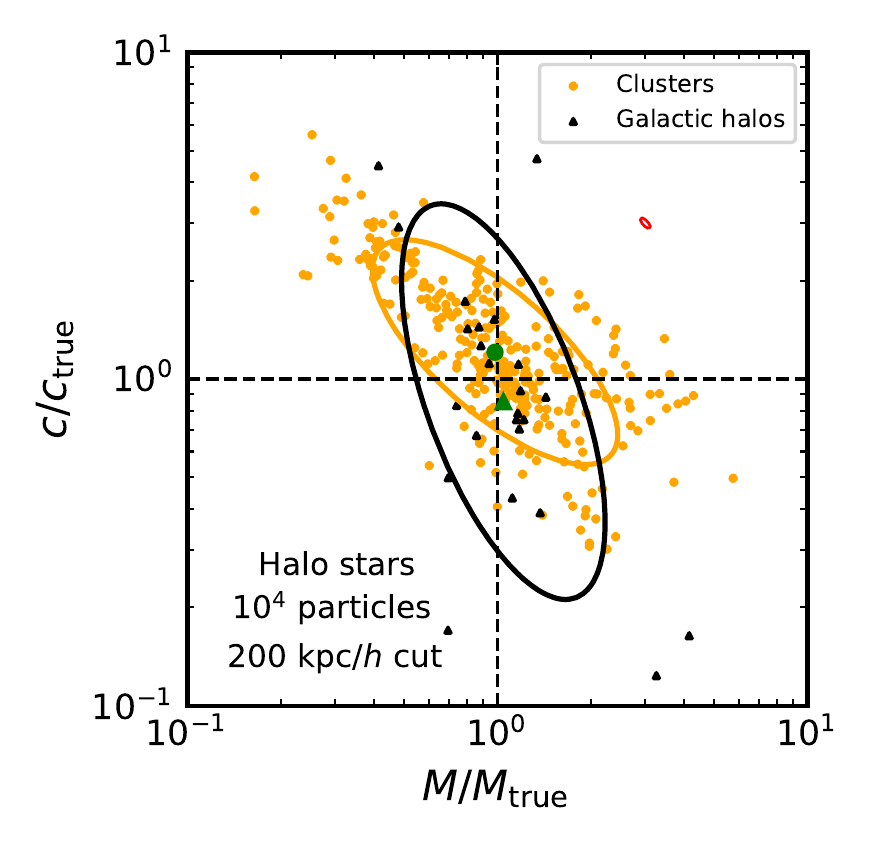}
    \caption{The best-fitting mass and concentration parameters of cluster and galactic haloes using star particles as tracers. For cluster haloes, star particles with $200\ h^{-1}{\rm kpc}<r<r_{200}$ and are not bound to any satellite galaxies are used as tracers. Each orange point shows the best-fitting result of one cluster halo, while each black point shows that of a single galactic halo taken from \citet{Wang2017} with an inner radius cut of 20~kpc. The orange and black ellipses represent the 1$\sigma$ scatters of the measurements, respectively. The red ellipse shows the averaged size of statistical uncertainty derived from the likelihood functions.}
    \label{fig:overall_star}
\end{figure}

Fig.~\ref{fig:overall_star} shows the best-fitting mass and concentration parameters of our cluster haloes using intracluster star particles as tracers. For each halo, we randomly select $10^4$ star particles located between 200 $h^{-1}\rm kpc$ and $r_{200}$ as tracers. Particles that belong to satellites are excluded. The mean bias and amount of scatters are provided in Table~\ref{tab:all}. The results are also compared with those from \citet{Wang2017} who modelled 24 galactic haloes from the APOSTLE simulation \citep{Fattahi2016,Sawala2016}, which is a set of zoomed-in hydrodynamical simulations of paired galaxies like our MW and M31. For each galaxy used by \citet{Wang2017}, $\sim 10^4$ to $\sim 10^5$ star particles in 20~kpc $<r<r_{200}$ that are not bound to any satellites were used as tracers. The typical statistical error is shown by the red ellipse in the upper right corner, which is again very small. As a result, the biases in the best-fitting parameters are primarily due to systematic errors as in the DM case. 

It appears that the average biases in the best-fitting parameters differ between the two samples of haloes. However, these differences are not significant given the large halo-to-halo scatters in the fits. In particular, the galactic sample consist of only 24 halos with a large halo-to-halo scatter of $\sigma_{\log c/c_{\rm true}}=0.4$ in the individual biases. This translates to an uncertainty in their average bias as $\sigma_{<\log c/c_{\rm true}>}=0.4/\sqrt{24}\approx 0.08$, comparable to the net bias. According to this we are interpreting the fits as "ensemble unbiased".

However, different from the results using DM particles, cluster haloes show similar systematic biases compared with galactic haloes. This is likely the result of two competing effects. On one hand, cluster haloes have later formation times and are thus less relaxed. On the other hand, the intrinsic number of phase independent particles also depend on the relative contribution of particles from different satellites. As discussed in \citet{Wang2017} (see their Equation 9), a coarse estimate of the $N_{\rm eff}$ can be found as
\begin{equation}
    N_{\rm stream,eff}=\frac{(\sum n_i)^2}{\sum n_i^2},\label{eq:Nstream}
\end{equation} where $n_i$ is the number of tracer particles in a single "stream" that are expected to originate from the same progenitor. According to this, $N_{\rm eff}$ is expected to be large when different satellites contribute similar number of particles, while it is smaller when only a few satellites dominate the contribution to the stellar halo. Indeed, the satellite population of cluster haloes have a steeper stellar mass function ($\ud N/\ud \ln M_\ast \sim M_\ast^{-1.6}$) than galactic haloes ($\ud N/\ud \ln M_\ast \sim M_\ast^{-1.2}$, see \citealp{Yang2009}) at the low mass end. If we assume the satellites are stripped by similar amounts, this means the stellar halo of a cluster receives relatively more contribution from a large number of low mass satellites, while that of a galaxy receives relatively more contribution from fewer massive satellites (see also discussion in \citealt{Deason2021}). This difference in satellite contribution can help to lower the value of $N_{\rm eff}$ in galactic haloes, bringing them closer to those in cluster haloes. These two effects together may result in the comparable intrinsic number of phase-independent particles. Nevertheless, we also want to point out that the above estimate of scatter for galactic haloes is derived from a relatively small sample of only 24 haloes. In addition, the galactic haloes used by \citet{Wang2017} are all paired, whereas the cluster haloes used in this paper are isolated. As has been shown by \citet{Wang2017}, paired haloes tend to show larger scatters than isolated haloes. 

Comparing the fitting results using halo stars in Fig.~\ref{fig:overall_star} with those based on DM particles in Fig.~\ref{fig:overall_DM}, it is very clear that the systematic uncertainties using star particles as dynamical tracers is much larger for both galactic and cluster haloes. This can be understood as DM particles in subhaloes are stripped earlier than star particles in satellites, which stay in the very inner parts of subhaloes. Thus star particles, after getting stripped, have less time to reach the steady state, and are more correlated in phase-space than DM particles. The more concentrated distribution of stars than DM in their progenitor satellites also makes it more difficult to remove the tight phase correlations among stars. 
For clusters, the estimated intrinsic number of star particles is about $N_{\mathrm{int}} \sim 20$. 


\subsection{Satellite galaxies} \label{sec:sat}

\begin{figure}
    \centering
    \includegraphics[width=0.5\textwidth]{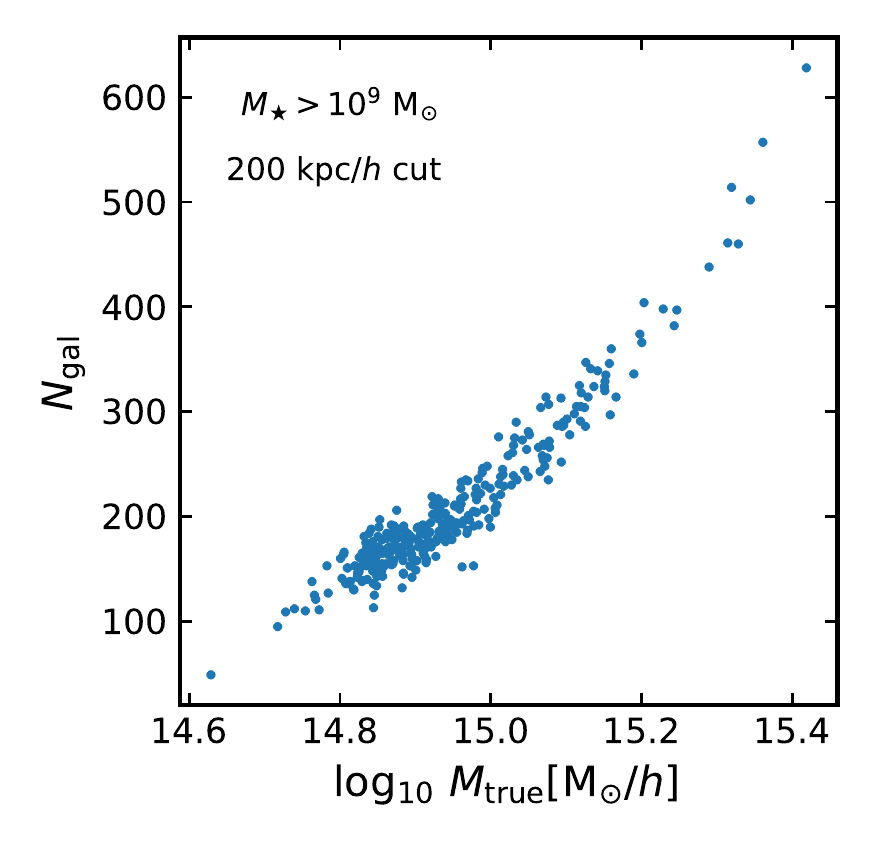}
    \caption{The number of satellite galaxies reported as a function of the host halo mass in our sample. Satellite galaxies with a stellar mass larger than $10^9\ \rm M_{\odot}$ and located between 200 $h^{-1}$kpc and $r_{200}$ from the halo centre are counted.}
    \label{fig:Ngal}
\end{figure}

\begin{figure}
    \centering
    \includegraphics[width=0.5\textwidth]{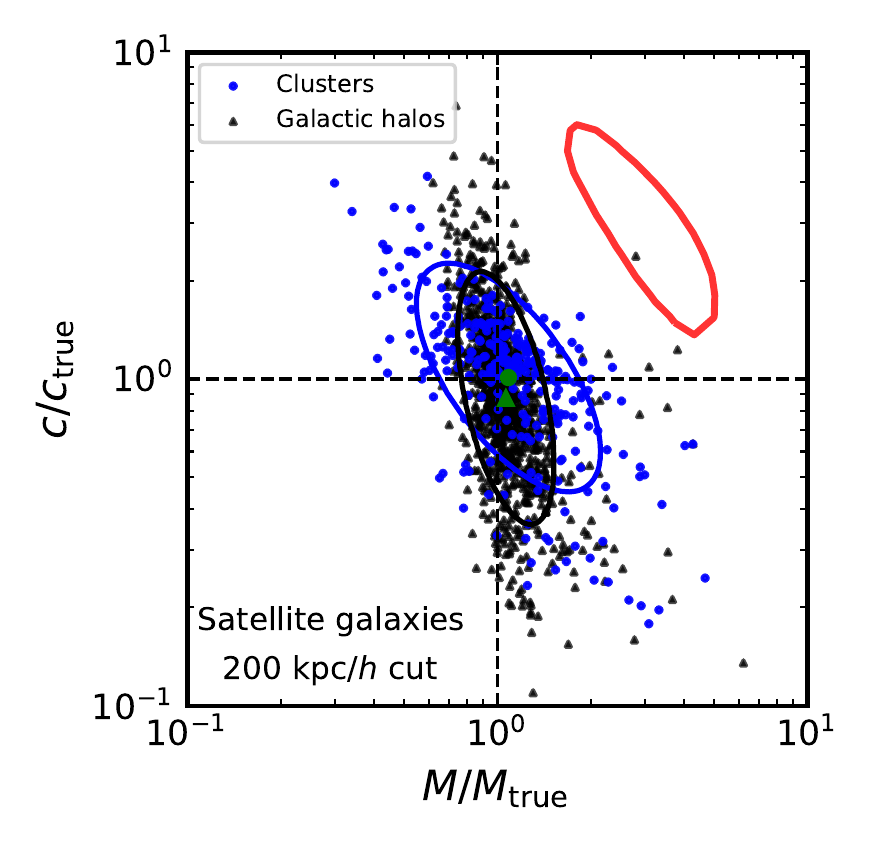}
    \caption{The best-fitting mass and concentration parameters using satellite galaxies as tracers. Each blue dot represents the fit to one cluster. Black triangles show the results based on galactic haloes taken from the Millennium-II simulation \citep{Boylan-Kolchin2009}, in which satellites located between 20 kpc and 400 kpc to the halo centre are selected as dynamical tracers (see text for details). The red ellipse in the upper right corner shows the average statistical error of individual clusters, while the blue and black ellipses indicate the 1$\sigma$ scatters of the best fits with the same colour. The green solid upper triangle and dot in the middle represent the mean biases of all galactic and cluster haloes, respectively, which are close to zero and almost ensemble unbiased.
    }
    \label{fig:overall_gal}
\end{figure}

Despite their sparsity, satellite galaxies are expected to better trace the dynamical state of DM~\citep{Han2020}. We select satellite galaxies with stellar masses larger than $10^9\ \rm M_\odot$ and a radial distance beyond 200 $h^{-1}\rm kpc$ from the halo centre.\footnote{We have repeated our analysis by choosing a mass cut of $>10^{10}\ h^{-1} \rm M_{\odot}$ to select satellite galaxies, and the results are very similar.} The number of satellite galaxies in each halo after this selection is shown in Fig.~\ref{fig:Ngal}, with an average number of $\sim 200$. The best-fitting halo mass and concentration parameters for both cluster and galactic haloes using satellite galaxies as dynamical tracers are shown in Figure~\ref{fig:overall_gal}. The satellite galaxies in isolated galactic haloes \footnote{Here isolated haloes refer to those without a massive neighbour, in contrast to binary systems like the MW-M31 pair.} are identified with the subhalo finder \citep[HBT+,][]{Han2018} from the Millennium-II 
simulation \citep{Boylan-Kolchin2009} in the radial range of $20-400\ \mathrm{kpc}$. The average number of all satellites in the galactic halo sample is about $\sim 500$.\footnote{The number of tracers is larger than that used in \citet{Han2020} due to the larger radial range adopted and due to the use of \textsc{HBT+}, which is able to identify more subhaloes.} The best fits are almost ensemble unbiased, and the scatter of galactic haloes is smaller than that of cluster haloes in mass estimates. The same conclusion holds also for their systematical uncertainties when we subtract the statistical errors from the total scatter. This is consistent with the expectation that galactic haloes are closer to steady states given their earlier formation times. For both galactic and cluster haloes, the halo mass and concentration parameters are anti-correlated. However, the exact directions for such anti-correlations are different between galactic and cluster haloes. As have been discussed in \cite{Wang2017}, the direction of anti-correlations between halo mass and concentration parameters is determined by the half-mass radius of tracers. The different correlation directions between galactic and cluster haloes thus mostly reflect the different radial distributions (including different radial ranges) of tracers in the two samples.

The mean mass estimate with respect to true values is 1.21 with a scatter of 0.20 dex. This is consistent with the results of \citet{Old2014,Old2015}, who compared the performances of more than 20 different cluster mass estimators using mock satellites populated with halo-occupation-distribution models and a semi-analytical galaxy formation model. They reported typical dispersions of $0.2-0.5$ dex in the recovered to true mass ratio for most methods. Our result is also comparable to the dispersion of 0.15 dex found by \citet{Saro2013} with a 3-D velocity dispersion based mass estimator.

Given the small number of satellite galaxies as tracers, the statistical uncertainty is no longer negligible. However, the total amount of scatter in the best-fitting parameters is still larger than the averaged statistical errors, revealing the existence of systematics, which are likely caused by the phase correlations due to the group infall of satellite galaxies. In the $\Lambda$CDM cosmology, haloes merge hierarchically, leading to the formation of satellite groups that share similar orbits around the host halo centre initially~\citep{HBT}. Using the Eq.~\eqref{eq:Nint}, we roughly gauge a systematic error of $\sim 0.13$ dex in mass with $N_{\rm int} \sim 3\times 10^2$.

\subsection{Comparisons of different tracers}

\begin{figure}
    \centering
    \includegraphics[width=0.5\textwidth]{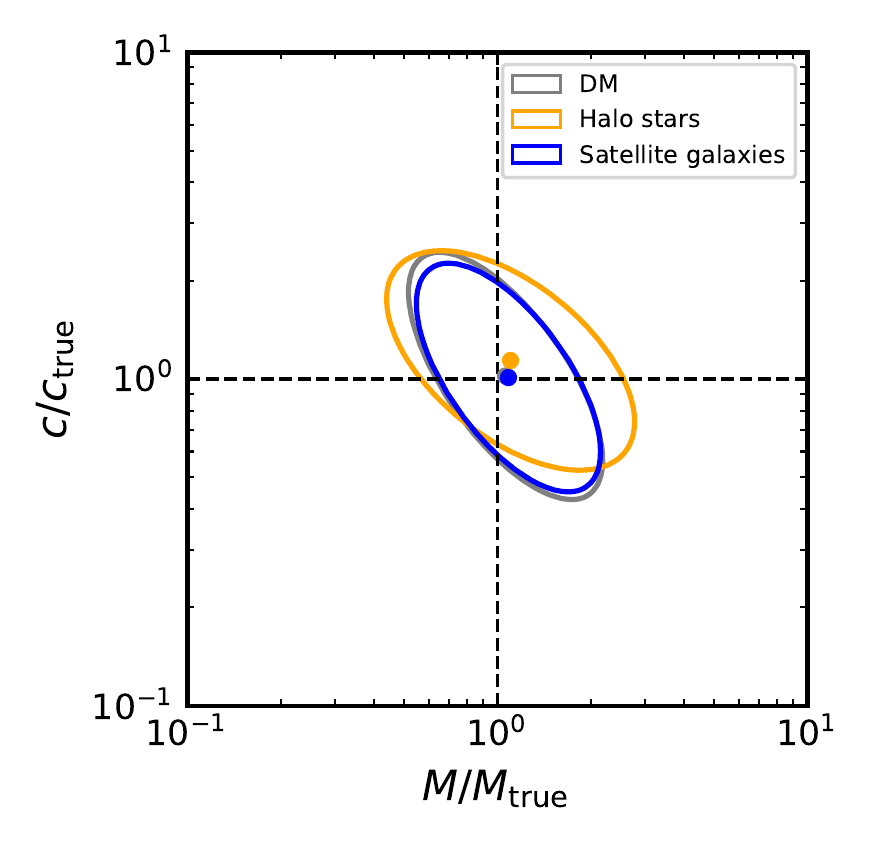}
    \caption{A comparison of the 1-$\sigma$ scatters in the best-fitting halo mass and concentration parameters, when different types of tracers are used. The grey, orange and blue points indicate the mean values using DM, halo stars and satellite galaxies as tracers respectively, while the ellipses of the corresponding colours show the 1$\sigma$ scatters. The number of tracer particles used are the same for different types of tracers. The DM and satellite galaxies show similar amounts of scatters, while the scatter for halo stars is slightly larger.}
    \label{fig:tracers}
\end{figure}

To make quantitative comparisons among the dynamical state of different types of tracers, we repeat the analysis using the same number of tracer particles for different tracers. More specifically, we downsample the DM and halo star samples to have the same number of tracer particles as the number of satellites in each galaxy cluster. This leads to similar amounts of statistical uncertainties, so that the differences in their total scatters are primarily contributed by different levels of systematics. As shown in Fig.~\ref{fig:tracers}, DM and satellite galaxies show a similar total scatter in the best-fitting parameters, while that from star particles is a little bit larger. This implies satellite galaxies closely trace the dynamical state of DM, while halo stars are less relaxed than the other two. The same conclusion has been reached by \citet{Han2020} for galactic haloes. $N_{\rm int}$ from satellite galaxies ($\sim3\times 10^2$) appears to be slightly larger than the one from DM ($\sim2\times 10^2$) as shown in Table~\ref{tab:all}, which we avoid over-interpreting due to the large statistical error of satellites.

The better performance of satellites compared with halo stars can also be understood according to Equation~\eqref{eq:Nstream}. Because we expect star formation to happen only within galaxies and we have excluded the very inner region corresponding to the central galaxy, the halo star population will be dominated by stars stripped from satellite galaxies. In a toy model, if we assume all particles stripped from the same satellite are completely correlated and only contribute to one independent phase-space measurement, then the effective number can be, at most, the number of satellites. Correspondingly, in Equation~\eqref{eq:Nstream}, $N_{\rm stream}$ is maximized when each stream contributes an equal number of tracer particles, and this maximum number equals the actual number of contributing streams. In the case of stellar halo, this means that the effective sample size of halo stars, at most, equals the number of contributing satellites, and the maximum number can only be reached when each satellite contributes equally to the stellar halo. In other words, the optimal weighting for halo stars is to weight different satellites equally, which corresponds to the case of using satellites as tracers. Note in the above arguments we did not consider that some satellite could have been completely disrupted, making the number of streams to be larger than the number of surviving satellites. Taking this into consideration could bring the $N_{\rm stream}$ from halo stars closer to the number of surviving satellites. The above arguments also apply to DM particles as tracers. However, in the DM case the contributions from completely disrupted subhaloes~\citep{SubGen} and smoothly accreted particles~\citep{Wang2011} are more prevalent than in the star case, resulting in comparable values of $N_{\rm int}$.

Compared with Fig.~\ref{fig:overall_star}, we find that the overall scatter using halo stars barely changes when using only a subset of tracer particles in Fig.~\ref{fig:tracers}. This indicates that the total scatter is dominated by the systematic error rather than the statistic error, so that simply changing the sample size does not change the total uncertainties significantly. In other words, the tracer sample size is already saturated because the total number of tracer particles greatly exceeds the number of phase-independent particles ($\sim 20$), and an increase in the total sample size does not help to bring in extra phase-space information.


\section{Discussions on additional systematics} \label{sec:sys}

In the previous section we mainly interpret the systematics in terms of correlated phase-space structures which deviate from the steady-state assumption. As our analysis also assumes spherical symmetry, deviations from it can also contribute to the systematics. Furthermore, the radial range of the tracer sample and the implemented baryonic physics can also play a role in the systematics. We now discuss these additional sources of systematics in this section.

\subsection{Spherical symmetry} \label{sec:sph}
\begin{figure}
    \centering
    \includegraphics[width=0.5\textwidth]{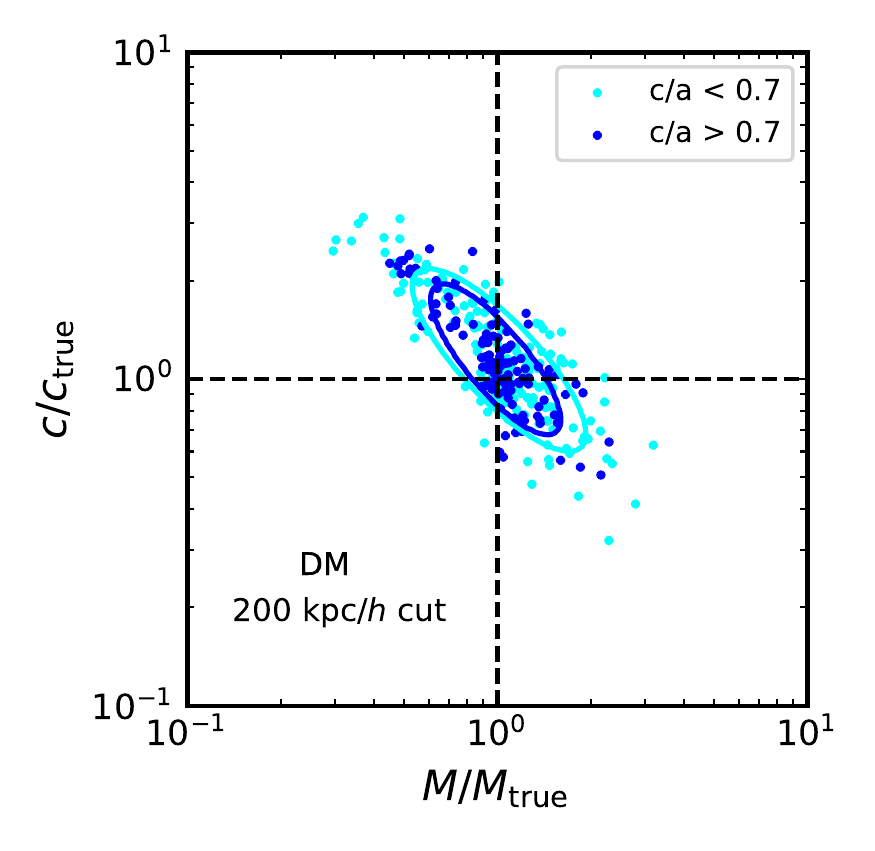}
    \caption{The distribution of best-fitting halo parameters for the two subsamples divided by the minor-to-major axis ratio, $c/a$, of the inertial tensor. The blue and cyan points show the best fits for haloes with $c/a > 0.7$ and $c/a < 0.7$, respectively, while the ellipses with the corresponding colours show the corresponding 1-$\sigma$ scatters.}
    \label{fig:axis}
\end{figure}

We use the minor-to-major axis ratio, $c/a$, of the eigenvalues of the inertial tensor given by the AHF haloes to mark the deviation from the spherical symmetry. To investigate the systematics caused by this deviation, we divide the sample into two equal subsets in $c/a$. In Fig.~\ref{fig:axis}, we compare the best-fitting results between the two subsets using \oPDF. 
Overall, the more spherical haloes ($c/a>0.7$) give a slightly smaller scatter as shown by the blue ellipse. A similar dependence was found by \citet{Wang2017} for isolated galactic haloes using DM particles as tracers. 

The difference between the two subsamples can be understood from two aspects. First, a smaller $c/a$ value means a stronger deviation from the spherical symmetry assumption of dynamical modelling, resulting in a larger systematic error. On the other hand, the shape of a halo also connects to its dynamical state, as they are both affected by the formation history and environment of the halo \citep[e.g.,][]{Morinaga2020,Gouin2021}. For example, a recent major merger can distort the halo shape and disturb the dynamical state simultaneously. In this sense, the halo shape can be viewed as a rough indicator of the dynamical state. As a result, part of the systematics that are correlated with $c/a$ can be attributed to deviations from the steady-state.

In a follow-up study (Li et al., in prep), we have carried out a comprehensive analysis to search for the most responsible variable for the total systematics. Here we briefly quote the relevant result -- the halo shape parameter is much less important compared to the other variables that characterise the dynamical state, such as the virial ratio to be defined in Section~\ref{sec:SJE}. Specifically, we quantify the level of systematics for each halo through a likelihood ratio variable, $\ln L_{\rm fit}/L_{\rm true}$, where $L_{\rm fit}$ and $L_{\rm true}$ are the likelihood values of the model evaluated with the best-fit and true parameters respectively. We find this likelihood ratio has a much larger correlation with the virial ratio than the axis ratio, indicating that the asphericity of haloes is a subdominant source of systematics.

\subsection{Radial Dependence}
Tracers at different radii of a halo may be subject to different levels of systematics, due to the variation of halo shape and the relaxation state with radius, for example. Observations may also impose radial selection effects in the tracer samples. For example, fibre collision and obscuration in real galaxy surveys can lead to incompleteness at very small scale. The real radius of a cluster is unknown before modelling, so the outer radial cut cannot be fixed at exactly the virial radius as we do by default in the previous analysis.
In this section, we investigate how the best-fitting results depend on the inner and outer radial cuts of the tracer samples, using DM, halo stars and satellite galaxies as tracers.

\begin{figure*}
    \centering
    \includegraphics[width=\textwidth]{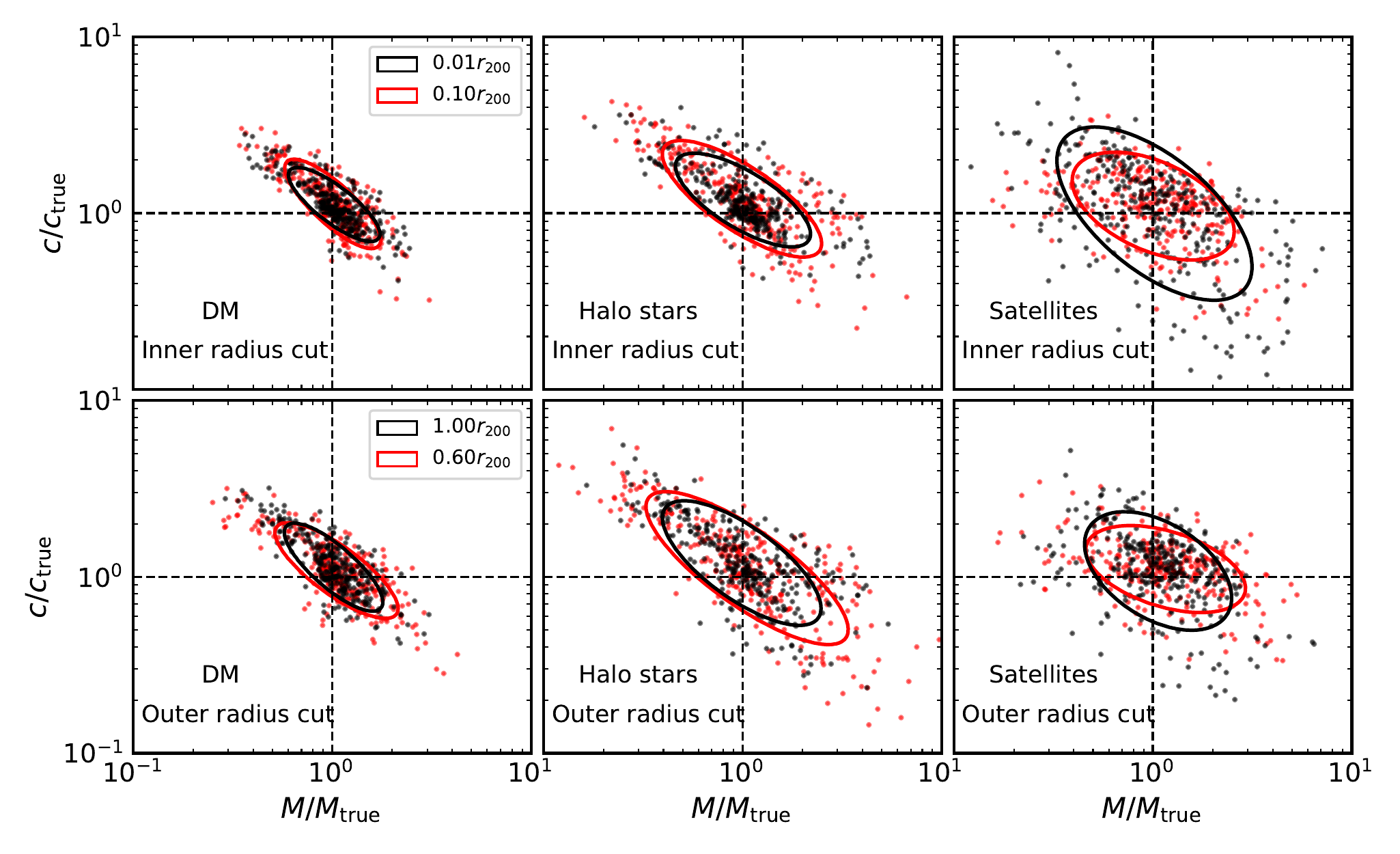}
    \caption{The best-fitting cluster mass and concentration parameters versus their true values, with varying inner (top panels) and outer (bottom panels) radius cuts in the tracer samples. The left, middle and right columns show the results using DM particles, halo stars and satellite galaxies as tracers, respectively. The points show the best fits to individual clusters, and the solid ellipses indicate their 1$\sigma$ scatters of the points, with different colours corresponding to different radial cuts as labelled. The numbers tracer particles used to fit each halo are $10^4$, $8\times10^3$ and 50 for DM, halo stars and satellites respectively.}
    \label{fig:rcut}
\end{figure*}

Fig.~\ref{fig:rcut} shows the best fitting results versus true halo parameters using various tracers with different inner and outer radial cuts as tracers. 
Throughout this paper, the default inner and outer radius cuts are $200\ h^{-1}\mathrm{kpc}$ and $r_{200}$, respectively. When varying the radius cuts, we limit the tracer sample sizes to be $10^4$ for DM particles, $8\times10^3$ for star particles and 50 for satellites in order to avoid running out of tracers for specific clusters.

In general, the total scatter decreases when the tracer sample extend to a smaller inner radius cut, especially when using DM particles as tracers. At the same time, the scatter also decreases with an increasing outer radial cut. These can be understood from two aspects. First, a larger radial range means a larger phase-space volume, leading to a larger number of phase-independent particles, thus decreasing the stochastic systematic uncertainty. Second, the wider radial range also means the overall shape of the potential profile can be better sampled \citep{Mamon2013}, leading to a better fit.


Compared with DM and halo stars, satellite galaxies do not present such a monotonic trend with the change in both inner and outer radius cuts. This can be likely interpreted as due to dynamical friction which sinks satellites to the halo centre, leading to stronger deviations from an equilibrium distribution in the inner halo~\citep[e.g.,][]{Old2013,SubGen}. 
Nevertheless, we avoid over interpreting these differences as they are also influenced by the large statistical uncertainties due to the very small number (50) of tracers.


\subsection{Baryonic effects} \label{sec:BE}

In this subsection, we further investigate how baryonic effects can affect our dynamical modelling results and whether they play an important role in changing the dynamical state of tracer objects in haloes. Throughout this paper, our default cluster sample is from the \gadgetx run -- one specific hydro-dynamically simulated galaxy clusters from the three-hundred project. To investigate the influence of baryonic effects on the dynamical state of clusters, we further apply our method to the same set of clusters but from its parent dark-matter-only run named MDPL2. As baryonic effects are expected to be important in the inner halo \citep[e.g.][]{Cui2014}, we avoid any inner radius cut and only impose an outer radius cut at the virial radius.

We use both the true potential template and the NFW profile in our modelling, in order to check whether the best-fitting halo mass and concentration are affected by how the underlying potential profile is modelled. In addition, as we mentioned before, it is impossible to know the true potential profiles in advance for real observed galaxy clusters. The potential templates adopted in previous sections are just helping us to understand the dynamical states of different types of tracers, while in this subsection we try the NFW parametrization which can be connected to real observations. 

\begin{figure}
    \centering
    \includegraphics[width=0.5\textwidth]{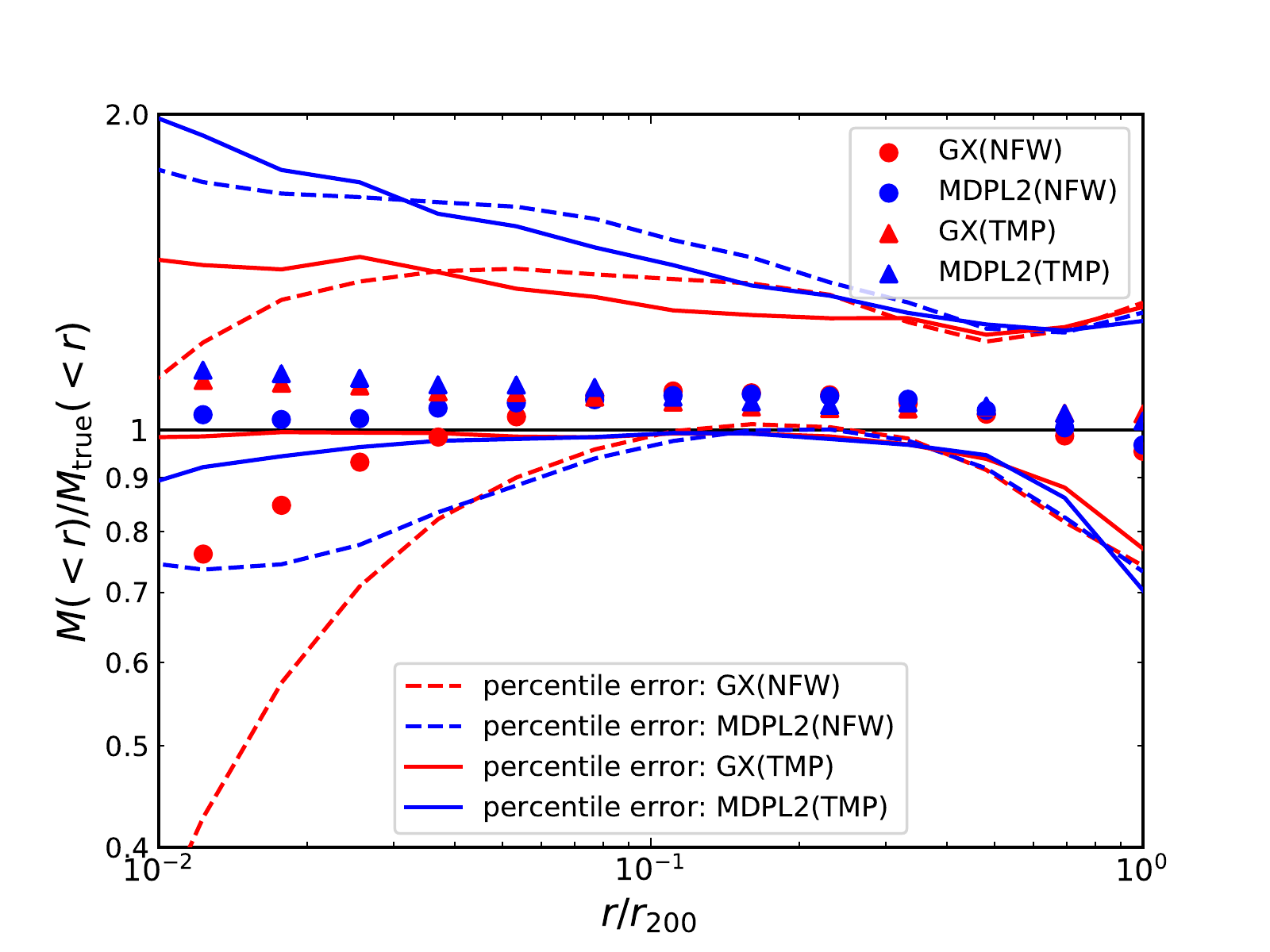}
    \caption{Distribution of the recovered halo mass profiles in the hydrodynamical (GX, shown with red coloured symbols) and dark-matter-only (MDPL2, blue colour) simulations. The data points are the median of the ratio between the best fit halo mass profiles and the true profiles, with filled circles showing the fits adopting NFW model profiles and triangles showing those using profile templates. The lines show the $16^{th}$ to $84^{th}$ percentile regions. Each cluster is fitted with the \oPDF method using $10^5$ smooth DM particles without any inner radius cut.}
    \label{fig:DMO_SPH}
\end{figure}

Overall, the amount of scatter in the recovered profiles are similar for all combinations of simulations and profiles as shown in Fig.~\ref{fig:DMO_SPH}, indicating similar dynamical states of the haloes in hydrodynamic and dark-matter-only simulations. However, significant net biases can be observed for the \gadgetx haloes in the inner part when modelled with the NFW profiles, while the same haloes show very little bias when modelled with profile templates. This means that the inner halo regions in the hydrodynamical simulations deviate more from the NFW form, consistent with previous findings that the inner density profile of haloes in hydrodynamical simulations are closer to an isothermal form instead of NFW~\citep[e.g.][]{Schaller2015,Cui2017}. 

A slightly smaller scatter is observed when modelling \gadgetx haloes with template profiles compared with MDPL2, especially in the inner part $r \lesssim 0.04 \times r_{200}$. To track down the reason, we have explicitly checked the axis ratio of the underlying matter distribution at different radii. We find that the matter distribution in hydrodynamically simulated galaxy clusters are slightly more spherical in the inner regions, compared to the dark-matter-only simulations, which might explain a smaller amount of scatter in the recovered inner mass profiles from hydrodynamical simulations.


The outer profile can be fitted with very small amounts of biases on average in all cases. This is because the outer halo is less affected by baryonic physics. Our results are also consistent with \citet{Armitage2019} who found no significant difference in the estimated halo masses between hydrodynamical simulations and dark-matter-only simulation. Note that the true halo mass difference between hydro and dark-matter-only simulated clusters is only several percents \citep[e.g.][]{Cui2012,Cui2014}.

\section{Generalization of the results} \label{sec:SJE}
Even though our analysis mainly focuses on the \oPDF method, most of the results are not limited to this method alone, as the steady-state and spherical symmetric assumptions in \oPDF are also frequently adopted by many other dynamical models. To demonstrate this, we repeat the analysis with the SJE ~\citep{Binney1987} applied to the smooth DM particles, as well as the generalized hydrodystatic equilibrium (GHE) equation applied to the intracluster gas. 

\subsection{Results from the Jeans equation}
The SJE, given by the first moment of the time-independent collisionless Boltzmann equation for spherically symmetric systems, is expressed as \citep{Binney1987}:
\begin{equation}
    \frac{\ud (v\sigma_r^2)}{v\ud r} + \frac{2\beta\sigma_r^2}{r} = -\frac{\ud \Phi}{\ud r} = -\frac{GM}{r^2},\label{eq:SJE}
\end{equation}
where $v$ is the number density profile of the tracer population, $\sigma_r$ is the radial velocity dispersion. The velocity anisotropy parameter, $\beta$, is defined as
\begin{equation}
    \beta \equiv 1 - \frac{\sigma^2_{\theta} + \sigma^2_{\phi}}{2\sigma_r^2},
\end{equation} where $\sigma_\theta$ and $\sigma_\phi$ are the velocity dispersions along the two tangential directions.

The left hand side of Eq.~\eqref{eq:SJE} can be measured from the distribution of tracer particles directly, thus can be used to infer the mass profile on the right hand side. The inferred mass profile from Eq.~\eqref{eq:SJE} is fitted with a template profile to estimate the halo parameters. When fitting the mass profile, we estimate the covariance matrix of different radial bins with a bootstrap method. Explicitly, we resample the tracer particles in each halo 200 times and repeat the analysis for each bootstrap sample. The estimated covariance from these bootstrap subsamples is then used to construct the chi square variable ($\chi^2$) for fitting the mass profile. The halo parameters are determined by minimizing the $\chi^2$ value using a $\textsc{python}$ package $\textsc{iminuit}$ with the minimiser $\textsc{minuit}$ based on \citet{James1975}. 

\begin{figure}
    \centering
    \includegraphics[width=0.5\textwidth]{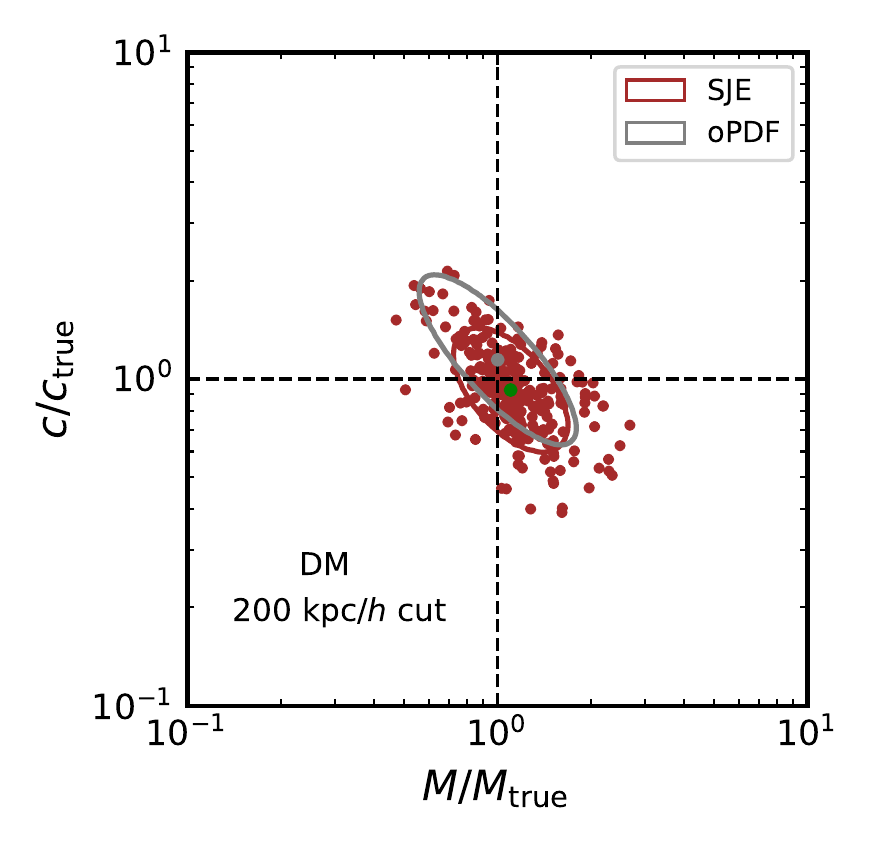}
    \caption{Best-fitting halo parameters normalised by their true values using smooth DM particles as tracers with two different methods, SJE and \oPDF. Each brown point represents the fit to one cluster using the SJE. Both SJE and \oPDF\ use $10^5$ tracers for each cluster. The scatter based on the SJE is shown by the brown ellipse, to be compared with the scatter of the \oPDF\ in Fig.~\ref{fig:overall_DM} reproduced here as the grey ellipse. The grey and green filled circles show the average values for the \oPDF and SJE fits respectively. The mean biases and overall scatters are provided in Table~\ref{tab:all}. The underlying potential profiles are modelled with a template form in the SJE case.}
    \label{fig:SJE}
\end{figure}

We present the results for the SJE method with smooth DM particles as tracers in Fig.~\ref{fig:SJE}. The number of tracers for each cluster is $10^5$ selected within the radial range from 200 $h^{-1}\rm kpc$ out to $r_{200}$. As the SJE method is also derived with the spherical and steady-state (or time independence) assumption, the results from \oPDF modelling are expected to be applicable to the SJE modelling. Overall, the SJE results show a comparable distribution of biases to that using the \oPDF, consistent with our expectation. More quantitatively, however, the SJE shows a slightly smaller total scatter than \oPDF. This can be understood as reflecting the different efficiencies of the two estimators, which we will discuss in more detail in section~\ref{sec:robustness}. 

To further demonstrate that the results from the \oPDF analysis can serve as dynamical benchmarks for different populations of haloes, we split our halo sample into relaxed and unrelaxed ones, and repeat the comparison between \oPDF and SJE. Following~\citet{Cui2018}, we make use of three variables to classify the dynamical state of haloes, including the virial ratio, $\eta$, the centre-of-mass offset, $\Delta_{\rm r}$, and the fraction of mass in subhaloes, $f_{\rm s}$ \citep[see][for example]{Cui2017}. The virial ratio is defined as $\eta\equiv(2T - E_{\mathrm{s}})/|W|$, where $T$ is the total kinetic energy, $E_{\rm s}$ is a surface pressure correction term and $W$ is the total potential energy. The centre-of-mass offset is defined as $\Delta_{\mathrm{r}}\equiv |\vec{r}_{\rm CoM} - \vec{r}_{\rm den}|/r_{200}$, where $\vec{r}_{\rm CoM}$ is the centre-of-mass location within $r_{200}$, and $\vec{r}_{\rm den}$ is the maximum density location of the halo. Haloes satisfying $0.85<\eta<1.15$, $\Delta_{\rm r}<0.04$ and $f_{\rm s}<0.1$ are classified as relaxed, while the remaining are classified as un-relaxed~\citep{Cui2018}.

\begin{figure*}
    \centering
    \includegraphics[width=\textwidth]{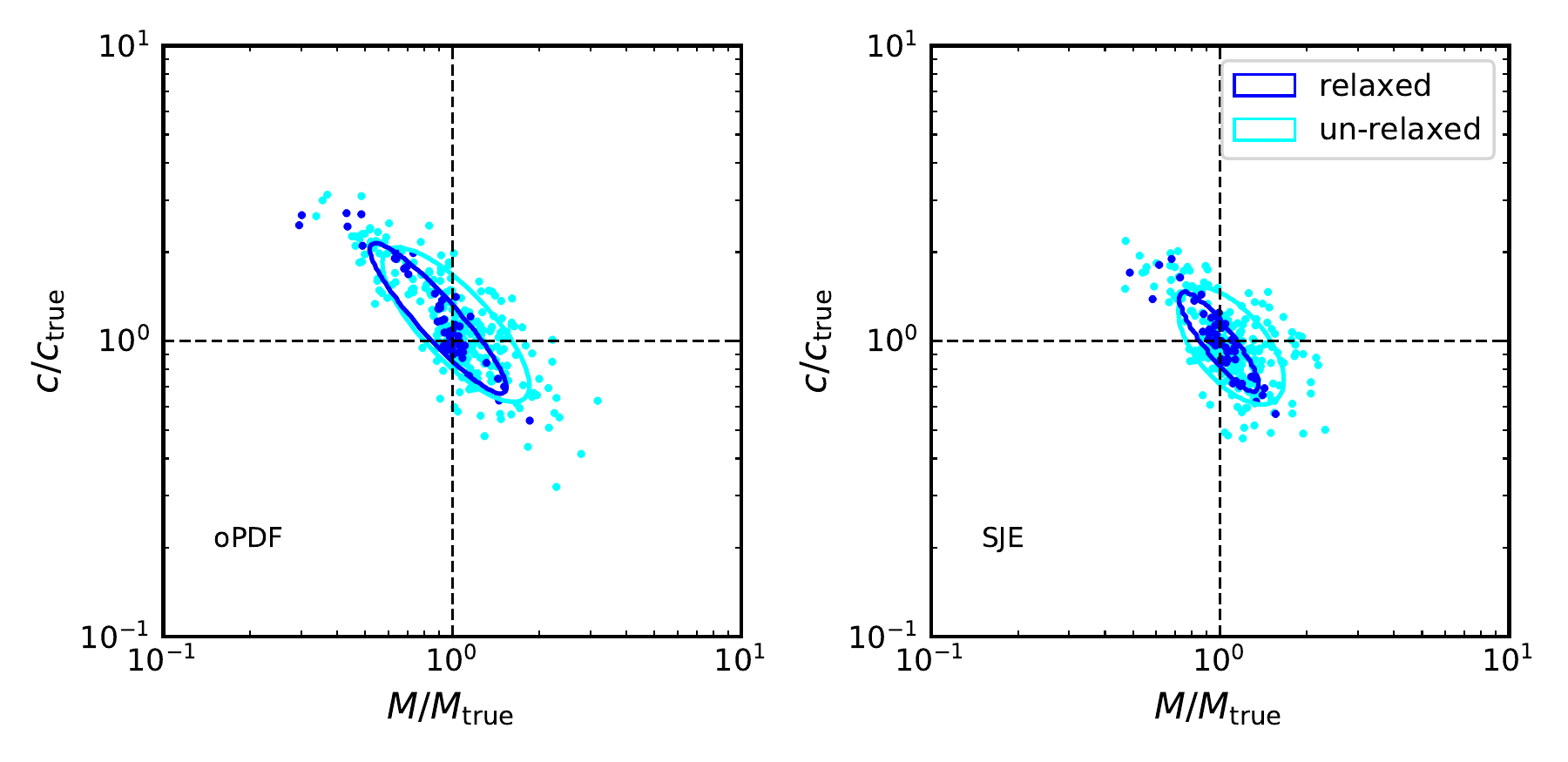}
    \caption{The distribution of best-fitting halo parameters for relaxed and un-relaxed clusters using smooth DM particles as tracers,. The blue and cyan points show the best fits for relaxed and un-relaxed clusters respectively, while the ellipses with corresponding colours show the 1-$\sigma$ scatters. The left and right panels are based on the \oPDF and the SJE modelling respectively.}
    \label{fig:DS}
\end{figure*}

In Fig.~\ref{fig:DS}, we compare the best-fitting halo parameters versus their true values for relaxed and un-relaxed clusters. The scatter of relaxed clusters is smaller compared to un-relaxed ones for both the \oPDF\ and SJE methods. This is because un-relaxed clusters are expected to deviate more from the steady state, and potentially also deviate more from spherical symmetry due to the lack of relaxation. 

Besides the SJE itself, we have also tried the virial theorem with a surface pressure term as used in \citet{Foex2017}, which can be derived from the SJE~\citep{Carlberg1997,Girardi1998,Armitage2018}. Overall, the systematics are similar to the SJE except for a net bias of $\log M/M_{\rm true}\sim -0.1$ which has also been reported in previous works~\citep{Foex2017, Armitage2018}. However, \citealp{Biviano2006} reported little net bias of ${M/M_{\rm true}}=1.03$ using a sample of $\sim 60$ clusters. The difference is likely due to the exact implementation of the virial theorem in different works~\citep{Armitage2018}. For the focus of the current paper, we choose to not further explore these differences here. 

\subsection{Intracluster gas modelled with hydrostatic equilibrium}
The SJE is only applicable to collisionless tracers. An extension to the Jeans equation with the presence of gas pressure is the generalised hydrostatic equilibrium (GHE) equation, which is normally used as a mass estimator in X-ray observation. For a spherical and steady system of gas, the balance between gravity and the pressure and velocity support reads~\citep{Rasia2004}:

\begin{equation} \label{HE1}
    \frac{\ud\Phi(r)}{\ud r} = \frac{1}{\rho(r)}\frac{\ud P(r)}{\ud r} + \frac{1}{\rho(r)}\frac{\ud (\rho(r) \sigma_r^2(r))}{\ud r} + \frac{2 \beta(r) \sigma_r^2(r)}{r},
\end{equation}
where $\rho$ is the mass density at radius $r$ and $\sigma_r$ is the radial velocity dispersion. 
This equation differs from the classical hydrostatic equilibrium in the last two terms to the right which describe the velocity support from the gas bulk motions.
Treating ICM as an ideal gas, the pressure can be expressed as $P(r)=\rho(r) k_\mathrm{B} T(r) / \mu m_\mathrm{p}$. This leads to the hydrostatic equilibrium mass estimator as
\begin{align} \label{HE2}
     M(<r) =&- \frac{r k_\mathrm{B} T(r)}{G \mu m_\mathrm{p}} \left[\frac{\ud \ln{\rho(r)}}{\ud\ln{r}} + \frac{
    \ud\ln{T(r)}}{\ud\ln{r}}\right] \\\nonumber
    & - \frac{r \sigma_r^2(r)}{G} \left[\frac{\ud\ln{\rho(r)}}{\ud\ln{r}} + \frac{
    \ud\ln{\sigma_r^2(r)}}{\ud\ln{r}} + 2\beta(r)\right], 
\end{align}
which we use to model the cluster mass profile from the distribution of gas particles.

To be consistent with the X-ray observations, we select hot gas particles as tracers with a temperature, $T > 0.3\ \rm keV$, and gas density, $\rho < 0.1\ \rm cm^{-3}$ (i.e. lower than the star-forming threshold). Same as in the \oPDF\ analysis, we randomly select $10^5$ smooth hot gas particles as tracers in the radial range of $200\ h^{-1}\rm kpc$ to $r_{200}$ for each cluster. Note that when applying Eq.~\eqref{HE2}, all the quantities on the right hand side, including the anisotropy and temperature profiles, are measured directly from the distribution of the smooth hot gas particles. In real observations, however, the anisotropy profile is often not available, which can lead to additional uncertainties or biases beside what we find below (see \citealp{Wang2018} for detailed discussions).

\begin{figure}
    \centering
    \includegraphics[width=0.5\textwidth]{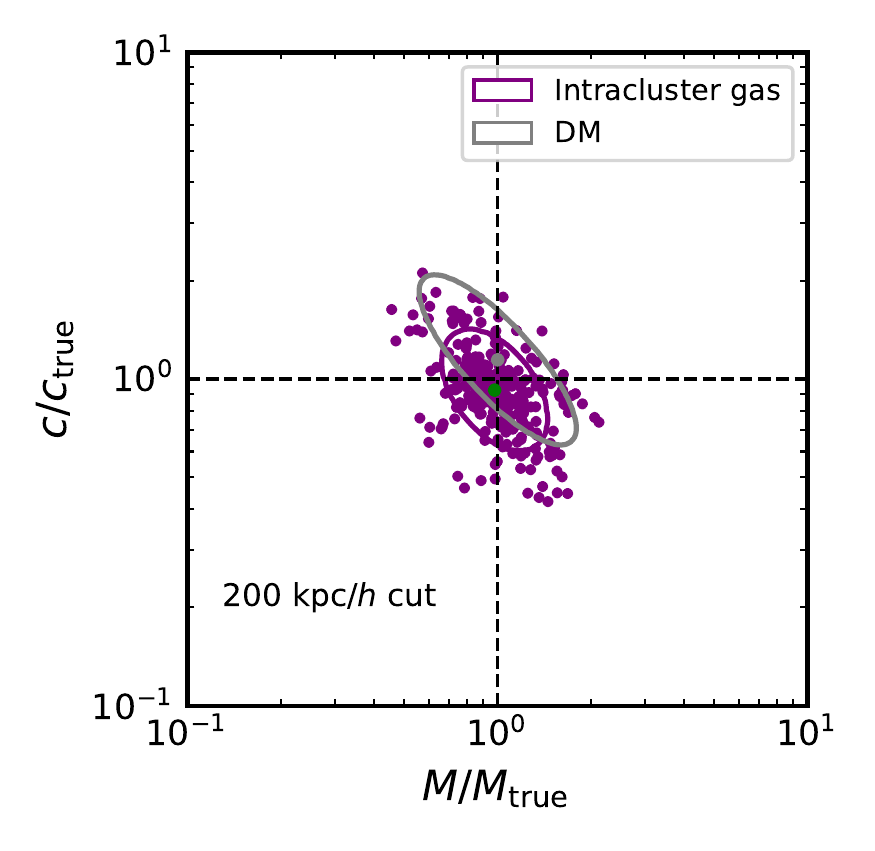}
    \caption{The best-fitting halo mass and concentration parameters of cluster haloes using smooth hot gas particles as tracers. The mass profile derived from the generalised hydrostatic equilibrium equation Eq.~\eqref{HE2} is fitted with a template profile to infer the halo parameters. Each point shows the fit to one cluster, and the purple ellipse shows the $1\sigma$ scatter of the points around their mean value marked by the green circle. For comparison, the grey ellipse shows the scatter of the smooth DM fits as shown in Fig.~\ref{fig:overall_DM}.}
    \label{fig:overall_gas}
\end{figure}

The process to estimate halo parameters from GHE is the same as from SJE. The distribution of the best-fitting parameters are shown in Fig.~\ref{fig:overall_gas}. Note that the statistical uncertainty in the GHE method is even smaller than in the DM case, thus not shown here. Similar to the results of the \oPDF fitting with other types of tracers, the deviations of the best-fitting halo mass and concentration parameters are anti-correlated and vary stochastically from halo to halo. The similar pattern of correlations suggests that the best constrained part of the mass profile using gas is located at a similar radius as using DM and satellites. This is also consistent with the results in Fig.~\ref{fig:DMO_SPH}, where the baryonic effects are only influencing the parts well within the best-constrained radius. The size of total scatter from the gas fits is comparable to that using DM as tracers. The similar performances of the gas and DM tracers can be interpreted as gas is largely in equilibrium with the DM, so that the pressure term in the GHE largely traces the velocity dispersions of DM. 
Our results are consistent with \citet{Foex2017} who found very good agreements between the dynamical and hydrostatic mass estimates after the removal of substructures. 

\subsection{The robustness of dynamical estimators} \label{sec:robustness}
Because the results in both Fig.~\ref{fig:SJE} and Fig.~\ref{fig:overall_gas} are obtained using $10^5$ tracer particles, the amounts of statistical errors are all negligible. Therefore, the total scatters in the plots are close to the systematic uncertainties in each case. Given the comparable amounts of scatters using different methods, it is thus tempting to conclude that the systematic uncertainty in the \oPDF represents a minimum uncertainty associated with all steady state models. However, this is not strictly correct as the systematic uncertainty in the SJE is slightly smaller than that of the \oPDF as shown in Fig.~\ref{fig:SJE}. This can be understood according to Eq.~\eqref{eq:Nint}, in which the systematic uncertainty depends on both the intrinsic number of phase-independent particles and the statistical uncertainty. Because the statistical uncertainty is determined by the efficiency of the dynamical model, a more efficient estimator can still show a smaller systematic uncertainty for the same tracer sample according to Equation~\eqref{eq:Nint}. Compared with $\Sigma_{\rm sys}$, $N_{\rm int}$ has factored out the dependencies on the efficiency (i.e., $N_{\rm tracer}\Sigma_{\rm sta}$) of the estimator, making it the more appropriate quantity to use when comparing the results from different dynamical estimators on different tracer samples.

\begin{figure*}
    \centering
    \includegraphics[width=\textwidth]{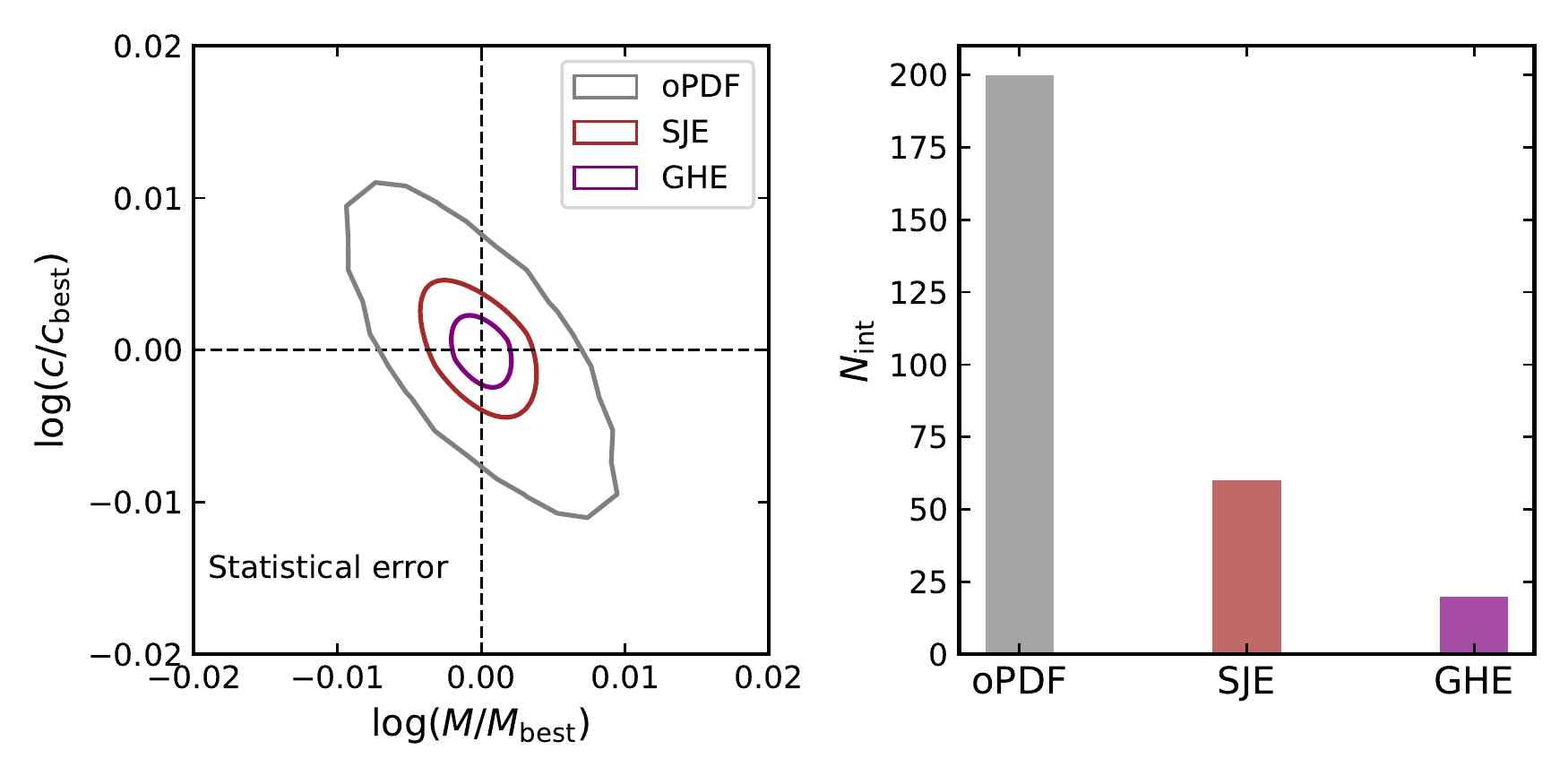}
    \caption{\emph{Left:} The average statistical errors for different dynamical models: \oPDF, SJE and GHE. The grey ellipse shows the statistical error for \oPDF when using $10^5$ DM particles as tracers corresponding to Fig.~\ref{fig:overall_DM}, while the brown and purple ellipses give the SJE and GHE results corresponding to the analysis in Fig.~\ref{fig:SJE} and Fig.~\ref{fig:overall_gas} respectively. 
    The centres of the statistical error ellipses are all translated to the origin to make a clear comparison. \emph{Right:} The intrinsic number of tracer particles $N_{\rm int}$ for \oPDF, SJE and GHE, calculated with Eq.\eqref{eq:Nint}. The values of $N_{\rm int}$ are presented at Table.~\ref{tab:all}.}
    \label{fig:stat_comp}
\end{figure*}

As shown in the left panel of Fig.~\ref{fig:stat_comp}, the statistical uncertainty in the SJE is smaller than that in the \oPDF. Combined with the total scatter, we infer $N_{\rm int}\approx 60$ using DM as tracers in the SJE, which is smaller than $N_{\rm int}\approx 2\times 10^2$ for \oPDF (see the right panel of Fig.~\ref{fig:stat_comp}). Naively, one would expect $N_{\rm int}$ from the SJE to be the same as that from \oPDF as they follow essentially the same set of assumptions. However, analogous to the varying statistical efficiencies, different estimators could also respond differently to the same sources of systematics. As a result, $N_{\rm int}$ can be understood as a measure of the \emph{robustness} of the estimator to systematics, with a larger $N_{\rm int}$ corresponding to a higher robustness because it corresponds to smaller $\Sigma_\mathrm{sys}$ according to equation~\eqref{eq:Nint}. In general, an optimal estimator is the one that minimizes the total uncertainty, $\Sigma_{\rm tot}$, for a given number of tracers, which depends on both the statistical efficiency and the robustness.

For the GHE fits to intracluster gas, the statistical uncertainty is much smaller than those of the \oPDF and SJE. The inferred intrinsic number of phase independent tracers is $N_{\rm int}\approx 20$ for gas, reflecting a low robustness of the GHE estimator. 
This can be interpreted as either reflecting the poorer equilibrium of the gas, or the presence of additional systematics in the GHE model that are degenerate with the phase correlations. 

It is encouraging to see that $N_{\rm int}$ is the highest for \oPDF. As additional systematics will enlarge the systematic uncertainty to reduce $N_{\rm int}$, we expect $N_{\rm int}$ inferred from \oPDF can broadly serve as an upper limit for  $N_{\rm int}$ associated with a general dynamical model for a given tracer type. Tracer samples with $N_{\rm tracer}>N^{\oPDF}_{\rm int}$ should have entered the systematics dominated regime for dynamical modelling.

\section{Conclusions} \label{sec:summary}
To study the limitations of dynamical models on estimating the cluster mass, we apply several representative dynamical methods to various types of tracers in 324 hydro-simulated galaxy clusters at $z=0$ from the Three Hundred project. We focus on testing how well the halo mass and concentration parameters can be recovered using the \oPDF~\citep{Han2016a} method, which is a generic minimal assumption dynamical model that relies only on the steady-state and spherical symmetry assumptions. The results are further validated using the hydrostatic equilibrium equation and the spherical Jeans equation.

In this work, we mainly focus on testing systematics due to deviations from the steady-state assumption. The underlying potentials of these galaxy clusters are modelled with parametric templates built from the true potential profiles, which enables us to bypass the systematics coming from the assumed functional form of the potential profile and to focus on investigating the steady-state assumption with different types of tracers. We further test the spherical symmetric assumption and the systematics when the underlying potential are modelled with the NFW form. In addition, we show the influences of adopting different inner and outer radius cuts of tracer particles on the modelling results, as well as the influences from baryonic physics by comparing clusters in hydrodynamical and dark-matter-only simulations.

\begin{table*}
    \centering
    \caption{Statistics on the distribution of best-fitting parameters. We list the mean values and the standard deviations for the biases in the best-fitting mass, $M/M_{\rm true}$, and concentration, $c/c_{\rm true}$ (abbreviated as $M_{\rm rat}$ and $c_{\rm rat}$ respectively). The correlation coefficient, $\rho_\mathrm{corr}$, between $\log M_{\rm rat}$ and $\log c_{\rm rat}$ is also listed for each case. The $\sigma(\log M_{\rm rat})$, $\sigma(\log c_{\rm rat})$ and $\rho_\mathrm{corr}$ value are used to derive the one-sigma ellipses in the corresponding figures. Note the quantities in brackets are the systematic error gauged from Eq.~\ref{eq:Nint}. The uncertainty of total error are estimated by bootstrapping. The $N_{\rm halo}$ column lists the actual number of haloes used in the analysis, while $N_{\rm int}$ lists the estimated intrinsic number of phase-independent tracers that would lead to the observed level of systematic bias (i.e., $\sigma$ values in the brackets). Note that different cluster samples have slightly different numbers of haloes because we have excluded a few catastrophic failures in each case. Also note the values of $N_{\rm halo}$ for different galactic samples are taken from \citet{Wang2017} for DM as tracers (MW-like haloes from Millennium-II), from \citet{Wang2017} for halo stars as tracers (MW-like galaxies from APOSTLE) and from \citet{Han2020} for satellites as tracers, which are thus not the same. The first part lists results from oPDF using template profiles while the second part lists those adopting either the GHE or the SJE method.}
    \label{tab:all}
    \begin{tabular}{c c c c c c c c c c c}
    \hline\hline
     halo & tracer & $N_{\rm halo}$ & $N_{\rm int}$ & <$M_{\rm rat}$> & <$\log M_{\rm rat}$> & <$\log c_{\rm rat}$> & $\sigma(\log M_{\rm rat})$ & $\sigma(\log c_{\rm rat})$ & $\rho_\mathrm{corr}$ & Fig.\\
     \hline
     \multicolumn{11}{c}{\oPDF}\\
     \hline
     Clusters & DM & 310 & $2\times10^2\pm17$ & 1.08 & 0.00 & $0.06$ & 0.17(0.17)$\pm 0.01$ & 0.17(0.17)$\pm 0.01$ & $-0.81$ & Fig.~\ref{fig:overall_DM} \\
     Galactic & DM & 1274 & $10^3\pm15$ & 1.02 & 0.00 & 0.01 & 0.08(0.08)$\pm 0.00$ & 0.13(0.13)$\pm 0.00$ & $-0.83$ & Fig.~\ref{fig:overall_DM} \\
     Clusters & Halo stars & 310 & $20\pm2$ & 1.17 & $-0.01$ & 0.08 & 0.26(0.26)$\pm 0.01$ & 0.23(0.23)$\pm 0.01$ & $-0.73$ & Fig.~\ref{fig:overall_star} \\
     Galactic & Halo stars & 24 & $40\pm13$ & 1.20 & 0.02 & $-0.07$ & 0.22(0.22)$\pm 0.04$ & 0.40(0.40)$\pm 0.06$ & $-0.61$ & Fig.~\ref{fig:overall_star} \\
     Clusters & Satellites & 304 & $3\times10^2\pm68$ & 1.21 & 0.04 & $0.00$ & 0.20(0.13)$\pm 0.01$ & 0.24(0.15)$\pm 0.01$ & $-0.65$ & Fig.~\ref{fig:overall_gal} \\
     Galactic & Satellites & 940 & $8\times 10^2\pm224$ & 1.10 & 0.03 & $-0.06$ & 0.10(0.06)$\pm 0.01$ & 0.26(0.16)$\pm 0.01$ & $-0.52$ & Fig.~\ref{fig:overall_gal} \\
    \hline
    \multicolumn{11}{c}{other dynamical models}\\
    \hline
     Clusters & DM (SJE) & 315 & $60\pm6$ & 1.15 & 0.04 & $-0.03$ & 0.12(0.12)$\pm0.01$ & 0.12(0.12)$\pm0.01$ & $-0.57$ & Fig.~\ref{fig:SJE}\\
     Clusters & gas (GHE) & 309 & $20\pm1$ & 1.01 & $-0.01$ & $-0.03$ & 0.11(0.11)$\pm0.01$ & 0.12(0.12)$\pm0.01$ & $-0.48$ & Fig.~\ref{fig:overall_gas} \\
    \hline
\end{tabular}
\end{table*}

The levels of systematics using various types of tracers and methods are summarised in Table~\ref{tab:all}. Our main conclusions are summarised as follows.

\begin{itemize}
    
\item The best-fitting halo mass and concentration parameters show large stochastic biases that vary from halo to halo, although their ensemble averages are almost unbiased, same as what has been found for MW-sized haloes~\citep{Wang2017,Wang2018}. This can be interpreted as mostly due to deviations from steady state as a result of phase correlations among tracer particles. Phase correlations lead to a reduced number of independent tracer particles than the actual number of tracers, resulting in a distribution of stochastic bias that behaves as an enlarged version of the statistical uncertainties. 

\item Galaxy clusters deviate more from steady state than MW-sized haloes, leading to larger stochastic biases. These amount to $\sim0.17$ dex in mass using smooth DM as tracers, compared to the $\sim 0.08$ dex biases for galactic haloes.


\item DM particles, halo stars and satellite galaxies have different amounts of phase correlations corresponding to different levels of systematic biases. Satellite galaxies and DM particles show similar amounts of systematic biases of $\sigma(\log M/M_{\rm true})\sim 0.13$ and 0.17 dex, indicating similar dynamical states, consistent with the results for galactic haloes~\citep{Han2020}. Halo stars show a higher bias level, revealing a higher amount of phase correlations as they are relatively difficult to strip from satellite galaxies and have less time to virialize compared with smooth dark matter particles, consistent with the results of \citet{Wang2017} for galactic haloes. 

\item Applying the generalised hydrostatic equilibrium equation, Eq.~\ref{HE2}, to the smooth hot gas particles results in a similar distribution of stochastic biases to that of DM particles using the \oPDF method, but with a smaller scatter of $\sim 0.11$ dex in mass.

\item The effective number of phase-independent tracer particles are about $2\times10^2$, $20$, $3\times10^2$ for DM, halo stars and satellites respectively. Dynamical modelling generally enters the systematics-dominated regime when the number of tracers is larger than these numbers, beyond which further increasing the tracer sample size can no longer significantly decrease the total uncertainty.

\item Particles in substructures/satellites can significantly worsen the dynamical modelling results when included in the tracer sample. This is because they strongly deviate from the steady-state assumption due to their clustering in phase space. In addition, substructure particles are also highly affected by the gravity from the substructure itself, besides that from the host halo. The influence of substructure particles is much stronger than what was observed in galactic haloes~\citep{Han2016b,Wang2017}, as cluster haloes are less relaxed and richer in substructures.

\item The performance of our dynamical modelling using \oPDF depends on both the inner and outer radial ranges of tracer particles. When the underlying potential is properly parametrized, enlarging the radial coverage of the tracer sample reduces the systematic biases. However, this is not true when the NFW profile is used to model the underlying potential in hydrodynamical simulations, where the baryonic effects cause the deviation from the NFW profile in the inner region ($r<\sim 0.3r_{200}$). Except for this very inner halo region, both the true potential templates extracted from the simulation and the NFW profile can reasonably recover the mass profiles, and using dark-matter-only or hydrodynamical simulations produce little difference in the halo mass bias. 


\item Our conclusions based on the \oPDF\ method have general implications for other dynamical modelling methods that also involve similar assumptions (steady state and spherical symmetry). We demonstrate this by comparing the \oPDF results against those from the spherical Jeans equation, which gives qualitatively similar results. The total uncertainty in the SJE modelling is slightly smaller than that in \oPDF, which is a result of the higher statistical efficiency but lower robustness in the SJE. 

This robustness of an estimator to systematics can also be quantified by the intrinsic number of phase independent tracers, $N_{\rm int}$. Comparing DM tracers modelled by different estimators, \oPDF exhibits the highest robustness with $N_{\rm int}\approx 2\times 10^2$, compared with $N_{\rm int}\approx 60$ for the SJE.
\end{itemize}

In this paper, we have mainly discussed the systematics caused by deviations from the steady-state assumption, and only briefly demonstrated that deviations from the spherical symmetry assumption is subdominant in the total systematics.
In a follow-up paper (Li et al., in preparation), we will further investigate several diagnostic variables of halo dynamical state including halo shape and the halo relaxation variables described in Section~\ref{sec:SJE}, to understand their independent contributions, as well as to find optimal proxies for separating haloes of different bias levels.

\section*{Acknowledgements}

The authors sincerely thank the anonymous referee for the careful and constructive comments, and Xiaokai Chen for the useful discussions.

This work is supported by NSFC (11973032, 11833005, 11890691, 11890692, 11621303, 12022307),
National Key Basic Research and Development Program of China
(No.2018YFA0404504) and 111 project No. B20019 and Shanghai Natural Science Foundation, grant No. 19ZR1466800. We gratefully acknowledge the support of the Key Laboratory for Particle Physics, Astrophysics and Cosmology, Ministry of Education. WC acknowledges supports from the European Research Council under grant number 670193 (the COSFORM project) and from the China Manned Space Program through its Space Application System. 

This work has been made possible by the ‘The Three Hundred’
collaboration. The project has received financial support from the European Union’s H2020 Marie Skłodowska-Curie Actions grant
number 734374, i.e. the LACEGAL project. 
The simulations used in this paper have been performed in the MareNostrum Supercomputer at the Barcelona Supercomputing Center, thanks to CPU time granted by the Red Espa$\tilde{\rm n}$ola de Supercomputaci$\acute{\rm o}$n. The CosmoSim database used in this paper is a service by the Leibniz-Institute for Astrophysics Potsdam (AIP). The MultiDark database was developed in cooperation with the Spanish MultiDark Consolider Project CSD2009-00064.

This work has made extensive use of the \textsc{python} packages -- \textsc{ipython} with its \textsc{jupyter} notebook \citep{ipython}, \textsc{numpy} \citep{NumPy} and \textsc{scipy} \citep{Scipya,Scipyb}. All the figures in this paper are plotted using the python matplotlib package \citep{Matplotlib}. This research has made use of NASA's Astrophysics Data System and the arXiv preprint server. The computation of this work is partly carried out on the \textsc{Gravity} supercomputer at the Department of Astronomy, Shanghai Jiao Tong University.

\section*{Data Availability}
The data underlying this paper will be shared on reasonable request to the corresponding author.



\bibliographystyle{mnras}
\bibliography{paper} 



\appendix

\bsp	
\label{lastpage}
\end{document}